\documentclass{article}%
\usepackage{amsmath}
\usepackage{amsfonts}
\usepackage{amssymb}
\usepackage{graphicx}
\usepackage{multirow}
\usepackage{fleqn}

\title{The fermionic King model}
\author{Pierre-Henri Chavanis, Mohammed Lemou, Florian M\'ehats }

\begin{document}

\maketitle{}

%\tableofcontents{}

\begin{abstract}
We study the fermionic King model which may provide a relevant model of dark
matter halos. The exclusion constraint can be due to quantum mechanics (for
fermions such as massive neutrinos) or to Lynden-Bell's statistics (for
collisionless systems undergoing violent relaxation). This
model has a finite mass. Furthermore, a statistical equilibrium state exists
for all accessible values of energy $E_{min}\le E\le 0$. Dwarf and
intermediate size halos are degenerate quantum objects stabilized against
gravitational collapse by the Pauli exclusion principle. Large halos at
sufficiently high energies are in a gaseous phase where quantum effects are
negligible. They are stabilized by thermal motion. Below a critical energy $E_c$
they undergo gravitational
collapse (gravothermal catastrophe). This may lead to the formation
of a central black hole that does not affect the structure of the halo. This may also lead  to the formation of a compact degenerate object surrounded by a hot massive atmosphere extending at large distances. We argue that large dark matter halos should not contain a
degenerate nucleus (fermion ball) because these nucleus-halo structures are
thermodynamically unstable. We compare the
rotation curves of the classical King model
to observations of large dark matter halos (Burkert profile). Because of
collisions and evaporation, the central density increases while the slope of
the halo density profile decreases until an instability takes place. We
find that the observations are compatible with a King profile at, or close to,
the point of marginal stability in the microcanonical ensemble. At that point,
the King profile can be fitted by the modified Hubble profile that has a flat
core and a
halo in which the density decreases as $r^{-3}$.  This is qualitatively similar
to the Burkert profile. Less steep halos are
unstable. Discrepancies between the King model and the observations are
interpreted as a result of incomplete relaxation.
\end{abstract}

\section{\label{intro}Introduction}
% References should be done using the \cite, \ref, and \label commands.
% Put \label in argument of \section for cross-referencing like this:
%\section{\label{}}

According to contemporary cosmology,  the universe is made of about $70\%$ dark
energy, $25\%$ dark matter, and $5\%$ baryonic (visible) matter \cite{bt}. Thus, the
overwhelming preponderance of matter and energy in the universe is believed to
be dark, {\it i.e.} unobservable by telescopes.  The dark energy is responsible
for the accelerated expansion of the universe. Its origin is mysterious and
presumably related to the cosmological constant or to some form of exotic fluid
with negative pressure such as the Chaplygin gas. On the other hand, dark matter
is necessary to account for the observed flat rotation curves of galaxies. Its
nature is one of the most important puzzles in particle physics and cosmology.

We consider the possibility that dark matter halos are described by the
Fermi-Dirac distribution at finite temperature. The
Fermi-Dirac distribution may
have two origins. First, it corresponds to the statistical equilibrium state of
a gas of fermions such as
massive neutrinos. These particles are subjected to the Pauli exclusion
principle similarly to electrons in white dwarf stars and neutrons in neutron
stars. As a result, quantum mechanics can stabilize dark matter halos against
gravitational collapse and lead to halo cores instead of $r^{-1}$ density cusps
predicted by the cold dark matter (CDM) model \cite{nfw} but not observed \cite{burkert}.
However, there is a
difficulty with this scenario because the time it takes a self-gravitating
system with a large number of particles to achieve a statistical equilibrium
state is
usually
very long and exceeds the age of the universe
by many orders of magnitude \cite{bt}.
Therefore, the establishment of a Fermi-Dirac distribution in the whole cluster is not granted. Furthermore,
the thermodynamical temperature is expected to be very low so that the cluster would be completely degenerate
and appear very different from what is observed (except in the case of dwarf halos).
Fortunately, there is another possibility. Dark matter halos may have reached a quasi stationary state (QSS) as a
result of a violent relaxation similar to the one imagined by H\'enon \cite{henonVR}, King \cite{kingVR} and
Lynden-Bell \cite{lb} for collisionless stellar systems such as elliptical
galaxies described by the Vlasov-Poisson
system. Coincidentally, the statistical prediction of Lynden-Bell \cite{lb} is also a
Fermi-Dirac-type distribution function, although the effective exclusion
principle is not due to quantum mechanics but to the Liouville theorem.
Furthermore, the temperature
appearing in Lynden-Bell's distribution is an effective
out-of-equilibrium temperature $T_{eff}$ that can be much larger than
the thermodynamical temperature. This could account for the value of the
temperature inferred from the rotation curves of the galaxies using the Virial
theorem. Finally, this collisionless relaxation occurs in a few dynamical
times and is much more efficient
than the collisional relaxation. In the dilute (non degenerate) limit, which is
appropriate to  large dark matter halos, the Fermi-Dirac
distribution reduces to the Boltzmann distribution.

If dark matter is collisionless, a halo should not evolve anymore after having
reached a virialized state. As a result, its central density cannot be very
high. In order to be more general, we consider the possibility
that the  core of dark matter halos is collisional \cite{spergel}.
This seems to be necessary to explain the presence of black holes at the center
of dark matter halos \cite{balberg}. When collisional effects are taken
into account, dark matter halos behave similarly to globular clusters. However,
the collisions between particles do not correspond to two-body encounters as in
globular clusters but rather to collisions similar to those in a
gas.\footnote{The
relaxation time due to strong short-range collisions is large (of the order of
the Hubble time) but, still, much smaller than the relaxation time due to weak
long-range encounters.} On the
other hand, in fermionic dark matter halos,  the Pauli exclusion principle must
be taken into account. As a result, collisions tend to establish a  Fermi-Dirac
distribution at finite temperature. They also allow the central concentration
of the system to evolve in time towards large values.

Self-gravitating systems have a very particular thermodynamics first investigated by Antonov \cite{antonov} and Lynden-Bell \& Wood \cite{lbw} in relation to
collisional stellar systems such as globular clusters made of classical point
mass stars. There is no statistical equilibrium state in a strict sense because
a self-gravitating system in an infinite domain has no entropy maximum (the
isothermal sphere corresponding to the Boltzmann distribution has infinite
mass) \cite{bt}. Therefore, the statistical mechanics
of self-gravitating systems is essentially an out-of-equilibrium problem. The
absence of statistical equilibrium state
is related to the fact that self-gravitating systems like globular clusters have the tendency to evaporate
(see, {\it e.g.}, Appendix A of Ref. \cite{sc2002}). This is already the case
for a non-interacting gas
if it is not enclosed within a container. However, evaporation is a slow
process and a globular cluster can be found, for intermediate times, in a quasi
stationary state close to the Michie-King distribution \cite{michie,king} which is a truncated
Boltzmann distribution with parameters slowly changing with time. If
we enclose the
system within a ``box'' so as to prevent artificially its evaporation
\cite{antonov,lbw}, it is found that
statistical equilibrium states exist only above a critical energy
$E_c=-0.335\, GM^2/R$ discovered by
Emden \cite{emden}. They have a density contrast $\rho(0)/\rho(R)<709$. These configurations are metastable
(local entropy maxima) but their lifetime is considerable since it scales as
$e^N$ (except close to the critical point) \cite{metastable}. For globular clusters, for which $N\sim 10^6$, this lifetime is so large that metastable states can be considered as stable states.
For $E<E_c$, there is no statistical equilibrium state.
The system
undergoes a gravothermal catastrophe \cite{lbw} and experiences core
collapse.

The evolution of a self-gravitating system is a three stages
process.\footnote{For a recent review on the kinetic theory of stellar systems,
emphasizing the pioneering works of Michel H\'enon to which this book is
dedicated, see Ref. \cite{aakin}.} (i) In the collisionless regime, a
self-gravitating
system initially out-of-equilibrium reaches a quasi stationary
state (virialized state) as a result of a  violent relaxation \cite{lb}.
(ii) In the collisional regime, the
system follows a sequence of King distributions \cite{michie,king} that are
long-lived metastable
equilibrium states \cite{metastable}. The evolution is driven by a
slow evaporation. During that stage, the halo expands while the core shrinks and
the central density increases as a consequence of the Virial theorem.  (iii) When the central
density reaches a critical value, the gravothermal catastrophe sets in and the
system undergoes core collapse \cite{cohn,lbe,heggie}. For classical self-gravitating systems, such as
globular clusters, core collapse leads to a binary star surrounded by a hot
halo.\footnote{This structure, ``binary star $+$ hot halo'', can be understood
simply in terms of thermodynamics. This is the most probable structure in the
microcanonical ensemble (MCE).
Indeed, we can increase indefinitely the entropy $S$ of a self-gravitating system at fixed mass and energy
by approaching two stars at very close distance and redistributing the released
energy in the halo in the form of kinetic energy (see, {\it e.g.}, Appendix A of
\cite{sc2002}). The binary
has a small mass $2m\ll M$ but a huge potential energy $E_{binary}\rightarrow -\infty$. Since the total energy $E$ is
fixed in MCE, the kinetic
energy (temperature) of the halo $T\rightarrow +\infty$ and, consequently,
the entropy $S\sim \frac{3}{2}Nk_B \ln T\rightarrow +\infty$. Since the
halo is ``hot'', it has the tendency to extend at large distances. It can be
shown \cite{sc2002} that the divergence of entropy is maximum when the mass in
the core is the smallest ({\it e.g.} a binary).} The formation of binaries at
the
center of globular clusters was anticipated
early by H\'enon \cite{henonbinaries,henonbinaries2}. These binaries can release sufficient energy to stop the collapse
and even drive a re-expansion of the cluster in a post-collapse regime \cite{inagaki}. This is followed
by a series of gravothermal oscillations \cite{oscillations,hr}. It is estimated that about $80\%$ of globular
clusters are described by the King model while $20\%$ have undergone core collapse. For self-gravitating
systems made of fermions (white dwarfs, neutron stars, dark matter halos) the
collapse stops when the core of the system becomes degenerate. This leads to a
core-halo structure with a
dense degenerate nucleus surrounded by a dilute atmosphere. In that case, we
can describe phase transitions between a gaseous phase unaffected by quantum mechanics and a condensed phase stabilized
by quantum mechanics. These phase transitions have been studied in detail by
Chavanis \cite{ijmpb} when the
system is confined within a box.

In the present paper, we make a step further and consider the fermionic King
model which is a truncated Fermi-Dirac distribution \cite{stella,mnras}. It can
be viewed as a generalization of the classical King model to the case of
fermions. The fermionic King model can be derived from a kinetic theory (based on the fermionic Landau
equation) by assuming that the particles escape the system when they reach a
maximum energy \cite{mnras}. This derivation is valid both for quantum particles (fermions)
and for collisionless self-gravitating systems undergoing Lynden-Bell's form of
violent relaxation.

The fermionic King model is interesting from the viewpoint of
statistical mechanics \cite{ijmpb}. Indeed, because of the energy truncation,
this distribution has a finite mass without the need to introduce an artificial
box. On the
other hand, because of the exclusion constraint (Pauli or Lynden-Bell bound),
there exist
a strict equilibrium state (global
entropy maximum) for all accessible values of  energy $E_{min}\le E\le 0$.
There may also
exist metastable
states (local entropy maxima) that can be as much, or even more, important than
fully stable states.\footnote{Indeed, the choice of the equilibrium state
depends
on a notion of ``basin of attraction'' and the metastable states may be reached more
easily from generic initial conditions than the fully stable states that
require very particular correlations. For example, in order to pass from the
gaseous phase
to the condensed phase the system must cross a huge barrier of entropy and
evolve through an intermediate phase in which some particles must approach very close to
each other. The probability of such a configuration
is extremely low \cite{metastable,ijmpb}.} As we
shall see, the nature of phase transitions in the
fermionic King model is similar to that already described in \cite{ijmpb} for
box-confined self-gravitating fermions. However, this model is more physical
(since there is no box) and it is therefore interesting to adapt the study of
\cite{ijmpb} to this more general situation.

The fermionic King model is also interesting from the viewpoint of astrophysics 
and cosmology because it provides a realistic model of dark matter halos. The
possibility that dark matter is made of fermions ({\it e.g.} massive neutrinos)
has been contemplated by several authors (see a detailed list of references in
\cite{cml}). Recently, de Vega  {\it et al.} \cite{vega3} have compared
fermionic models of dark matter halos with observations and obtained encouraging
results. However, they use the usual
Fermi-Dirac distribution. Since this distribution has infinite mass, it
is not fully realistic to describe dark matter halos. Furthermore, this precludes the possibility of
making a stability analysis and describing phase transitions.

In this contribution, we discuss the main properties of the
classical and fermionic King models in MCE. We plot the caloric curves and the
profiles of density and circular velocity. We study the stability of the
solutions by using the Poincar\'e theory of linear series of equilibria and
we describe phase
transitions between a gaseous phase and a condensed phase. We also compare our
theoretical predictions to the observations of dark matter halos. A more
detailed study is developed in  \cite{cml}.

\section{The classical and fermionic King models}
\label{sec_cf}

The fermionic King model is defined by the distribution function \cite{cml}:
\begin{equation}
f=A\frac{e^{-\beta(\epsilon-\epsilon_m)}-1}{1+\frac{A}{\eta_0} e^{-\beta(\epsilon-\epsilon_m)}} \quad {\rm if} \quad \epsilon\le \epsilon_m,
\label{cf1}
\end{equation}
\begin{equation}
f=0 \quad {\rm if} \quad \epsilon\ge\epsilon_m,
\label{cf2}
\end{equation}
where $f({\bf r},{\bf v})$ gives the mass density of
particles with position ${\bf r}$ and velocity ${\bf v}$,  $\rho({\bf r})=\int
f({\bf r},{\bf v})\, d{\bf v}$ gives the mass density of particles with position ${\bf r}$, $\Phi({\bf
r})$ is the gravitational potential determined by the Poisson
equation $\Delta\Phi=4\pi G\rho$, $\eta_0$ is the maximum accessible value of
the distribution function, $\epsilon=v^2/2+\Phi({\bf r})$ is the individual
energy of the particles, $\beta$ is the inverse temperature, $\epsilon_m$ is
the escape energy above which the particles are lost by the system, and
$\mu\equiv \eta_0/A$ is the degeneracy parameter.

For $\epsilon\ll \epsilon_m$, we can make the approximation
$e^{-\beta(\epsilon-\epsilon_m)}\gg 1$ and we recover the Fermi-Dirac
distribution $f=\eta_0/(1+e^{\beta\epsilon+\alpha})$ with $\alpha=\ln(\eta_0/A)-\beta\epsilon_m$ \cite{cml}. As recalled in the Introduction, the Fermi-Dirac distribution may
have two origins: (i) it describes a gas of fermions at statistical
equilibrium in which case $\eta_0=g m^4/h^3$ is the maximum accessible value of
the distribution function fixed by the Pauli exclusion principle ($h$ is the
Planck constant, $m$ the mass of the particles, and $g=2s+1$ the spin
multiplicity of the quantum states); (ii) it results from the violent
relaxation of a collisionless system of particles (classical or quantum) as
described by Lynden-Bell \cite{lb} and worked out by Chavanis and Sommeria
\cite{csmnras}.
In that case, Eqs. (\ref{cf1}) and (\ref{cf2}) are
valid for the coarse-grained distribution function (usually denoted
$\overline{f}$) and $\eta_0$ is the maximum value of the fine-grained
distribution function.  We shall consider the two possibilities since the
distributions are formally the same. In the quantum
interpretation $\beta={m}/{k_B T}$, where $T$ is the temperature. In
Lynden-Bell's interpretation $\beta={\eta_0}/{k_B T_{eff}}$, where $T_{eff}$ is
an effective (out-of-equilibrium) ``temperature''. In order to unify the notations,
we write $\beta=1/T$ where $T$ has the dimension of an energy.

The fermionic King model was introduced heuristically by Ruffini and Stella \cite{stella} as a natural
extension of the classical King model to fermions in order to describe dark
matter halos made of massive neutrinos. This distribution function was
independently introduced by Chavanis \cite{mnras}  where it was derived from a kinetic equation (the
fermionic Landau equation) assuming that the particles leave the system when
they reach a maximum energy $\epsilon_m$. The kinetic derivation given in \cite{mnras} is
valid either for quantum particles (fermions) or for collisionless
self-gravitating systems experiencing Lynden-Bell's type
of violent relaxation.

In the non degenerate limit $\mu\rightarrow +\infty$, we can make the approximation $\frac{A}{\eta_0}e^{-\beta(\epsilon-\epsilon_m)}\ll 1$ and we recover the classical King model
\begin{equation}
f=A \left\lbrack {e^{-\beta(\epsilon-\epsilon_m)}-1}\right\rbrack\quad {\rm if} \quad \epsilon\le \epsilon_m,
\label{cf3}
\end{equation}
\begin{equation}
f=0 \quad {\rm if} \quad \epsilon\ge\epsilon_m.
\label{cf4}
\end{equation}
For $\epsilon\ll \epsilon_m$, we can make the additional
approximation $e^{-\beta(\epsilon-\epsilon_m)}\gg 1$ and we recover the
Boltzmann distribution $f=\eta_0 e^{-(\beta\epsilon+\alpha)}$. The classical King model describes globular
clusters and, possibly, large dark matter halos for which degeneracy effects (due
to the Pauli exclusion principle for fermions or due to the Liouville
theorem for collisionless systems undergoing violent relaxation) are
negligible.

The fermionic King distribution  (\ref{cf1})-(\ref{cf2}) is a critical point
of the ``entropic'' functional
\begin{eqnarray}
S=&-&\int \biggl\lbrace  A \left \lbrack \left (1+\frac{f}{A}\right ) \ln \left
(1+\frac{f}{A}\right )-\frac{f}{A}\right\rbrack\nonumber\\
&+&\eta_0 \left \lbrack \left (1-\frac{f}{\eta_0}\right ) \ln \left
(1-\frac{f}{\eta_0}\right )+\frac{f}{\eta_0}\right\rbrack   \biggr\rbrace    \,
d{\bf r}d{\bf v}
\label{cf5}
\end{eqnarray}
at fixed energy $E=\frac{1}{2}\int f v^2\, d{\bf r}d{\bf v}+\int
\rho\Phi\, d{\bf r}$ and mass $M=\int \rho\, d{\bf r}$. Indeed, it
cancels the first order variations of the constrained entropy, $\delta S-\beta\delta E-\alpha\delta
M=0$, where $\beta$ and $\alpha$ are Lagrange multipliers associated with the conservation of energy and mass. In the non-degenerate
limit, the entropic functional associated with the classical King model
(\ref{cf3})-(\ref{cf4})  is
\begin{equation}
S=-\int A \left \lbrack \left (1+\frac{f}{A}\right ) \ln \left
(1+\frac{f}{A}\right )-\frac{f}{A}\right\rbrack\, d{\bf r}d{\bf v}.
\label{cf6}
\end{equation}
We shall be interested in entropy {\it maxima} at fixed energy and mass
\begin{equation}
S(E)=\max_f \lbrace S[f]\, |\, E[f]=E,\quad M[f]=M\rbrace.
\label{cf7}
\end{equation}
As discussed in \cite{katzking,cml}, this variational principle provides a condition of thermodynamical stability in MCE for tidally truncated self-gravitating systems. We
could also consider the canonical ensemble (CE) where the
temperature $T$ is fixed instead of the energy. In that case, the stable fermionic King distribution is
obtained by minimizing the free energy $F=E-TS$ at fixed mass. For
self-gravitating systems, the statistical ensembles are inequivalent (see,
{\it e.g.}, the reviews \cite{paddy,katzrevue,ijmpb}). In this
contribution, for conciseness, we restrict ourselves to MCE. This is the most
relevant ensemble for dark matter halos that are relatively  isolated objects
(their energy is approximately conserved).

Finally, we would like to make clear that we consider here the {\it thermodynamical} stability of the fermionic King
distribution. This implies that we are considering the stability
of the system with respect to a collisional evolution or with respect to a violent
relaxation on the coarse-grained scale (both described by the fermionic Boltzmann or Landau equation). We shall see that
the fermionic King distributions are not always thermodynamically stable, and this puts interesting
constraints on these distributions. On the other hand, we recall that the fermionic King distributions,
and more generally all the distribution functions of the form $f=f(\epsilon)$ with $f'(\epsilon)<0$, are nonlinearly
{\it dynamically} stable with respect to the Vlasov equation describing a
collisionless evolution  \cite{lmr}.

\section{Thermodynamics of the classical King model}
\label{sec_class}

We first consider the thermodynamics of the classical King model that
corresponds to the non-degenerate limit $\mu\rightarrow +\infty$. This distribution is appropriate to
describe globular clusters and it may also be appropriate to describe large dark matter halos.

\begin{figure}[!h]
\begin{center}
\includegraphics[clip,scale=0.3]{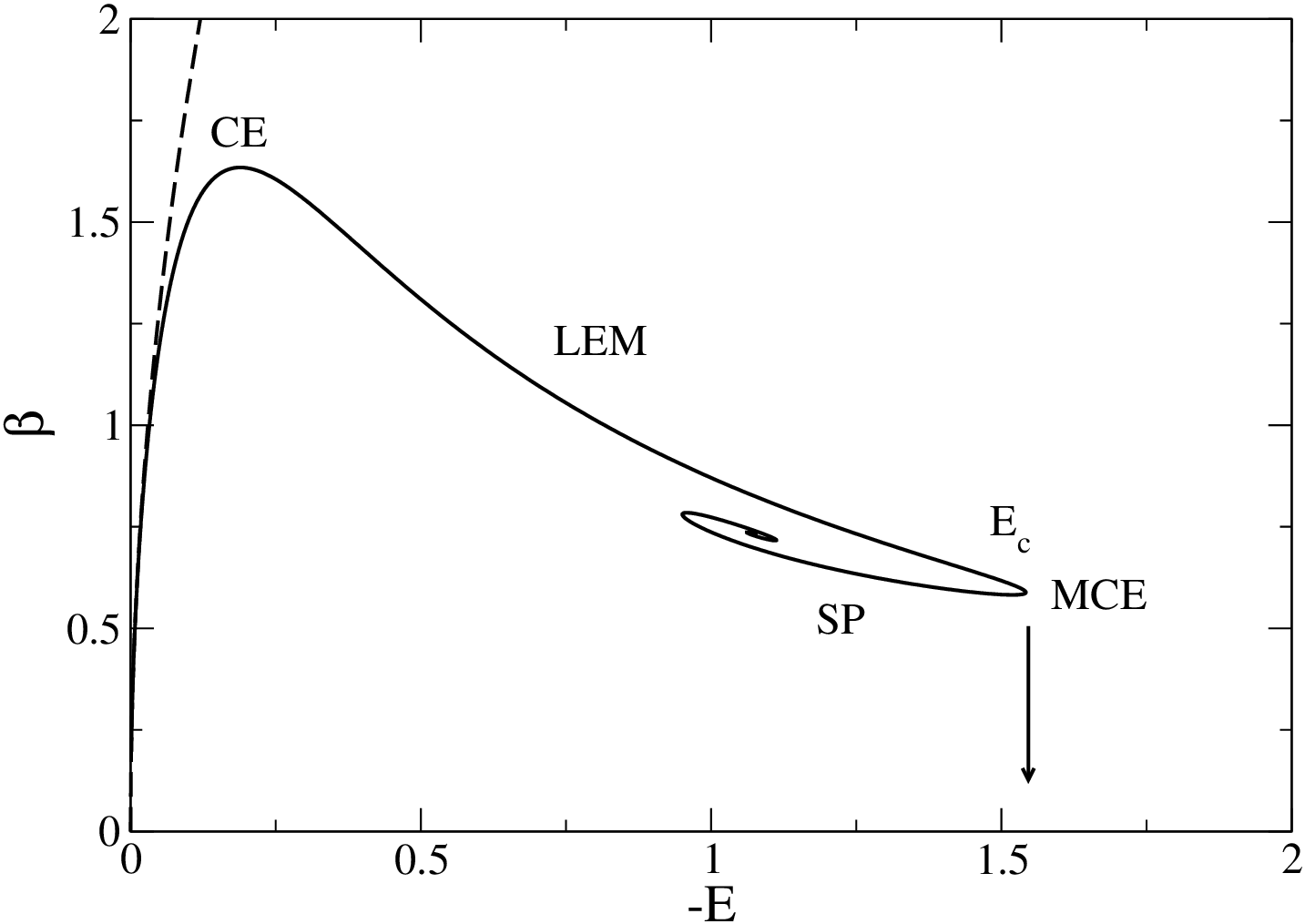}
\caption{Series of equilibria (parameterized by $k$) giving the inverse temperature $\beta$ as a function
of the energy $-E$ for the classical King model. The dashed line corresponds to the polytropic approximation.}
\label{ebAhenon}
\end{center}
\end{figure}

\begin{figure}[!h]
\begin{center}
\includegraphics[clip,scale=0.3]{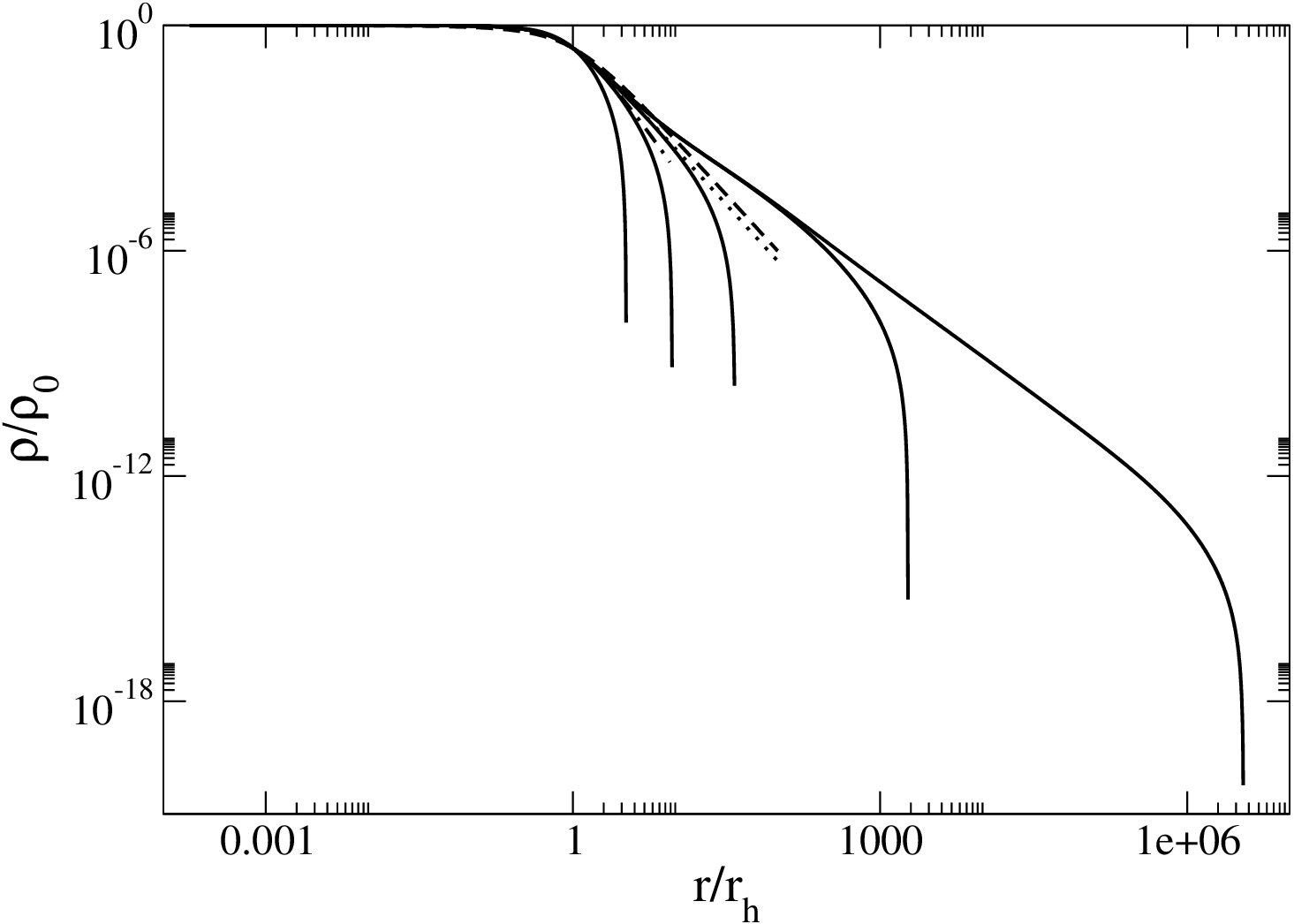}
\caption{Normalized density profiles of the
classical King model in logarithmic scales for (left to right): $k=1.34$
($E=-0.188$,
$\beta=1.63$), $k=5$ ($E=-0.965$, $\beta=0.893$), $k=7.44$ ($E=-1.54$,
$\beta=0.589$), $k=15$ ($E=-1.09$, $\beta=0.735$), and $k=30$
($E=-1.07$, $\beta=0.732$). We have defined the halo radius  $r_h$ such that
$\rho(r_h)/\rho_0=1/4$ \cite{vega3}. For $k\rightarrow +\infty$, the system approaches the
classical isothermal sphere and
the density decreases as $r^{-2}$ with damped oscillations superimposed \cite{chandra}. However, our study
shows that the King profiles with $k>k_{MCE}=7.44$ are thermodynamically unstable. Therefore, these oscillations may not
be physically relevant. As $k$ decreases, the effective slope of the density profile increases. For $k=k_{MCE}=7.44$
the density profile decreases approximately as $r^{-\alpha}$ with an effective
slope $\alpha\sim 3$. For $k=5$ the density profile
has an effective slope $\alpha\sim 4$. The King model is stable in MCE as long
as the
effective slope $\alpha$ is
approximately larger than $3$. The dotted line represents the modified Hubble profile which has a
slope $\alpha=3$ \cite{bt}. It fits
well the core of the isothermal sphere for $r<1.63 r_h$. It also fits well the
King model with $k\sim k_{MCE}$ up to
$\sim 5 r_h$. The dashed-dotted line represents H\'enon's isochrone profile that
has a slope $\alpha=4$ \cite{isochrone}. It fits
well the  King model with $k\sim 5$ up to $\sim 2 r_h$. The dashed line
represents the Burkert profile corresponding
to the observations of dark matter halos \cite{burkert}. It has a slope
$\alpha=3$. }
\label{densityLOG}
\end{center}
\end{figure}

\begin{figure}[!h]
\begin{center}
\includegraphics[clip,scale=0.3]{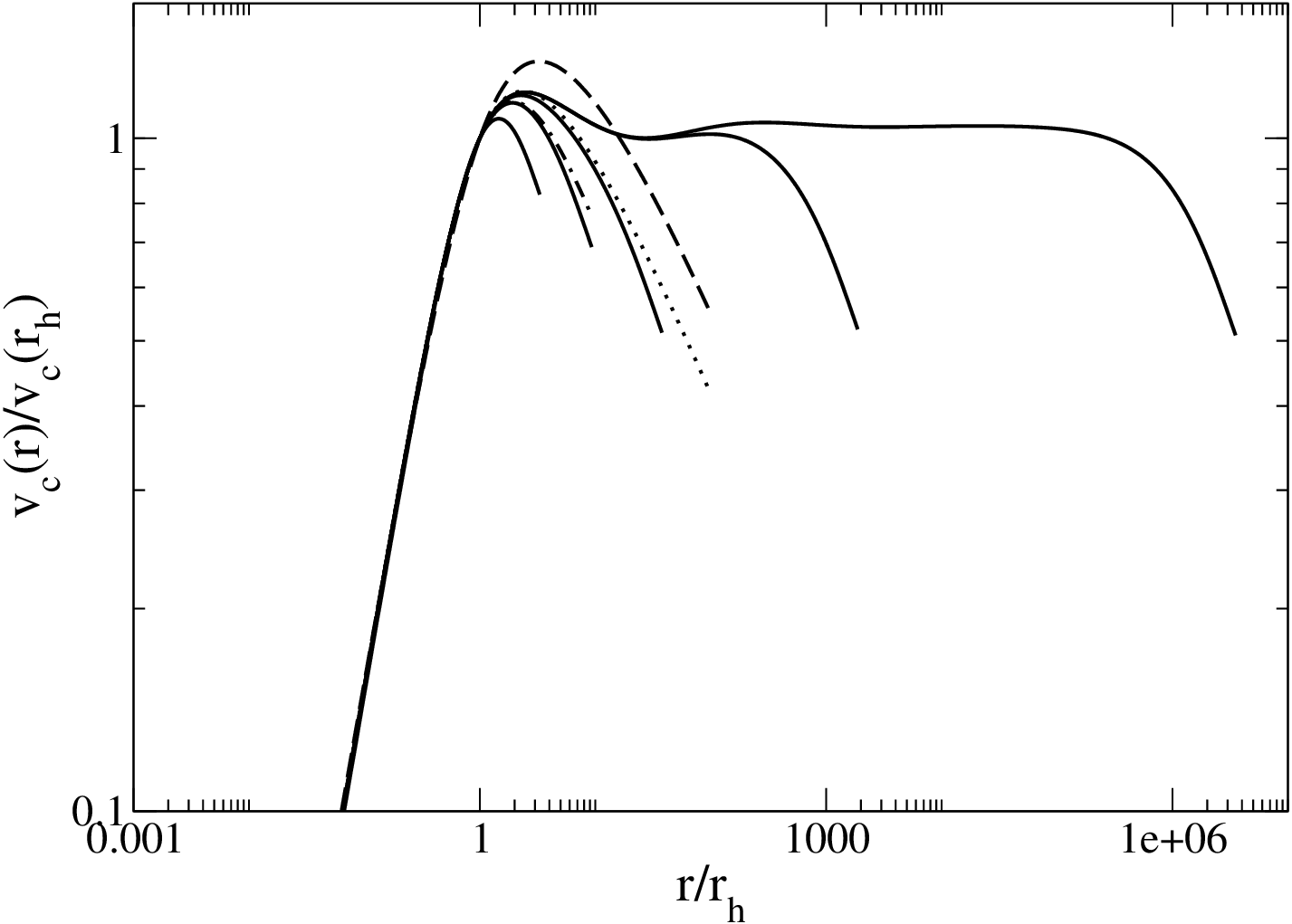}
\caption{Normalized rotation curves of the
classical King model in logarithmic scales for (left to right): $k=1.34$, $5$,
$7.44$, $15$,
and $30$.  For $k\rightarrow +\infty$, the system approaches
the classical isothermal sphere and the rotation curve presents a plateau with damped oscillations superimposed.
However, these oscillations occur at very large distances $r>100 r_h$
(probably not accessible observationally) and
our study shows that the King profiles with $k>k_{MCE}=7.44$ are unstable. Therefore, these oscillations
may not be physically relevant. For $k\sim k_{MCE}$, the rotation
curve presents a maximum close to the halo radius $r_h$ before decreasing, in
qualitative agreement with the observational Burkert profile (dashed line). The
modified Hubble profile (dotted line) and H\'enon's isochrone profile
(dashed-dotted line) provide a good fit of the King profiles with $k\sim
k_{MCE}$ and $k\sim 5$ respectively up to the tidal radius. }
\label{vitesseLOG}
\end{center}
\end{figure}

\begin{figure}[!h]
\begin{center}
\includegraphics[clip,scale=0.3]{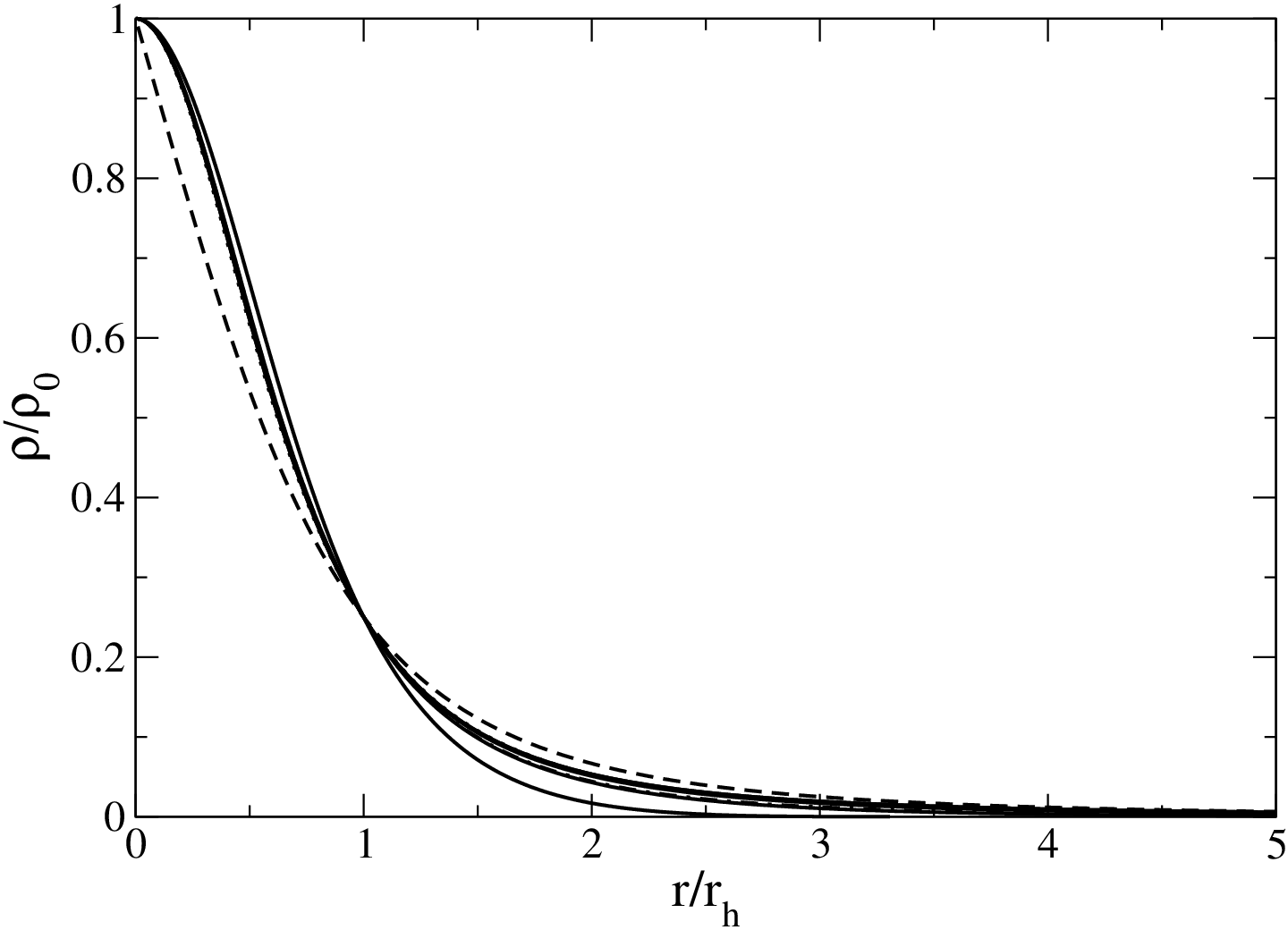}
\caption{Normalized density profiles of the
classical King model in linear scales
for (bottom to top): $k=1.34$, $5$, $7.44$, $15$, and $30$.}
\label{densityLIN}
\end{center}
\end{figure}

\begin{figure}[!h]
\begin{center}
\includegraphics[clip,scale=0.3]{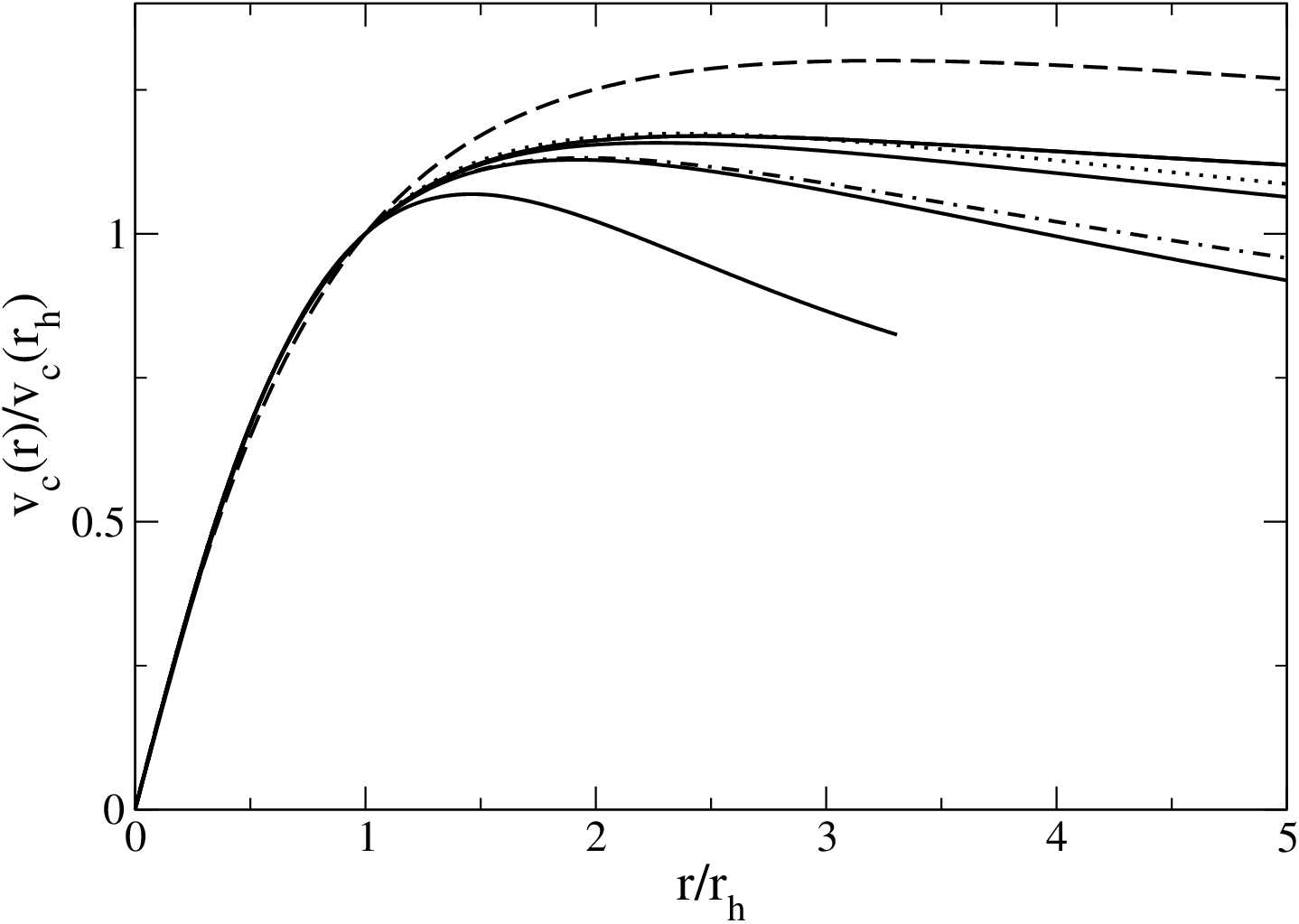}
\caption{Normalized rotation curves of the
classical King model in linear scales for (bottom to top): $k=1.34$, $5$,
$7.44$, $15$, and $30$.
}
\label{vitesseLIN}
\end{center}
\end{figure}

The series of equilibria  $\beta(-E)$ giving the inverse temperature
normalized by the quantity $1/[G^2M^{4/3}(8\pi\sqrt{2}A)^{2/3}]$ as a function
of the energy
normalized by the quantity $G^2M^{7/3}(8\pi\sqrt{2}A)^{2/3}$ is plotted in
Fig. \ref{ebAhenon}.\footnote{Some authors \cite{lbw,nardini} use another
normalization that is discussed in \cite{cml}.} It is parameterized by the
concentration parameter $k=\beta [\epsilon_m-\Phi(0)]$ going from $0$ to
$+\infty$ (see \cite{cml} for details). This
curve updates the one drawn by Katz \cite{katzking}. It has
a snail-like structure (spiral) similar to the series of equilibria of classical
isothermal spheres confined within a box (see, {\it e.g.},
\cite{ijmpb}).\footnote{We note,
however, that the energy is
always negative in the present case. This is a consequence of the Virial theorem
 for self-confined systems. By contrast, for box-confined
isothermal spheres, there is an additional term in the Virial theorem that
accounts for the pressure of the system against the boundary. As a result, the
energy may be positive.}
The concentration parameter $k$ increases along the series of equilibria. For small $k$, the
system is equivalent to a polytrope of index $n=5/2$ \cite{cml}. This approximation is
valid for $E\rightarrow 0^-$ and $T\rightarrow +\infty$. For large $k$, the
system is similar to the isothermal sphere ($n=+\infty$) and the series of
equilibria spirals about the limit point
$(E_{\infty},\beta_{\infty})=(-1.07,0.731)$. Some density profiles, and the
corresponding rotation curves $v_c(r)=\sqrt{GM(r)/r}$,
are plotted in Figs. \ref{densityLOG}-\ref{vitesseLIN}.

\begin{figure}[!h]
\begin{center}
\includegraphics[clip,scale=0.3]{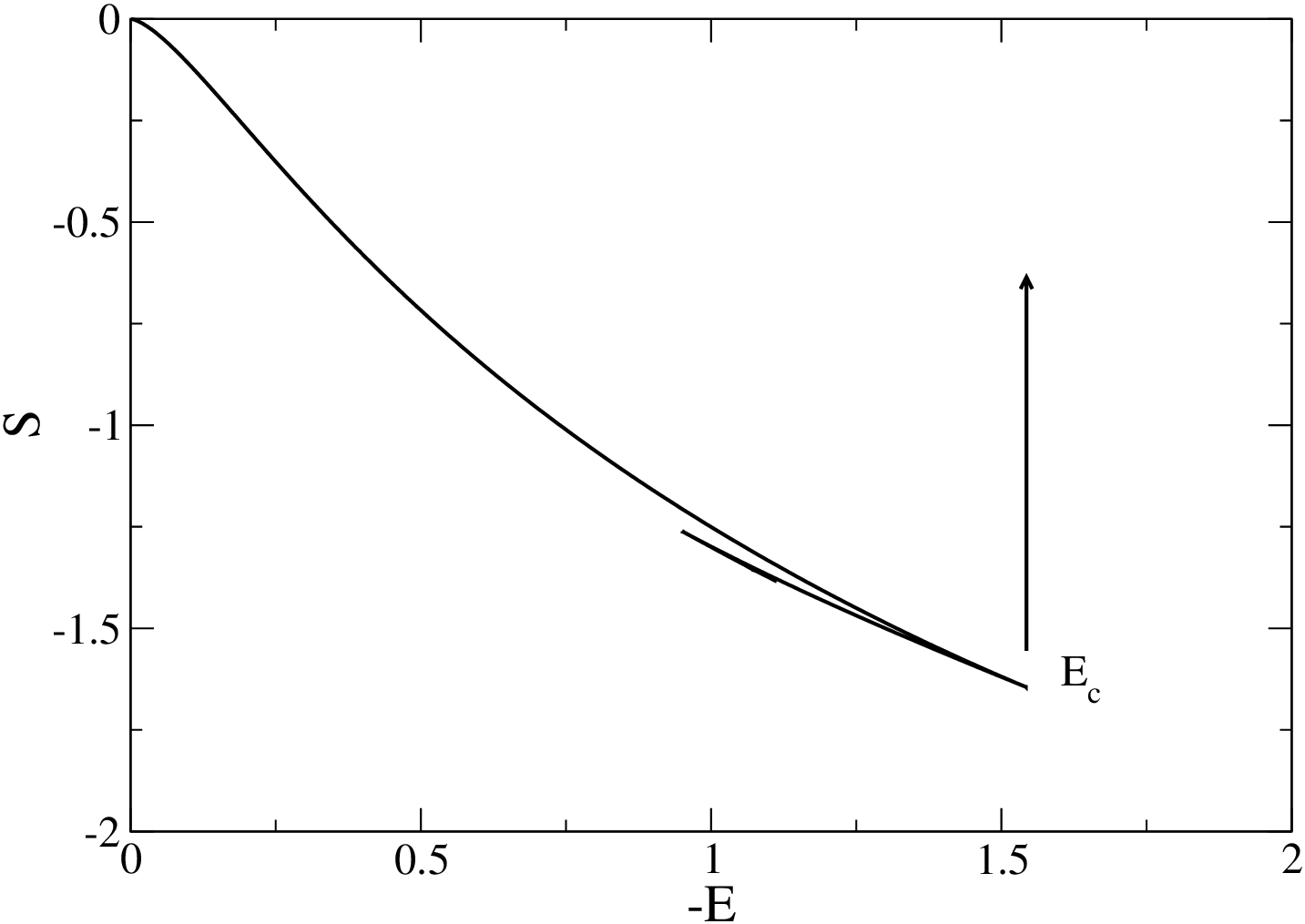}
\caption{Entropy versus energy for the classical King model.}
\label{esAhenon}
\end{center}
\end{figure}

In MCE, there exist equilibrium
states only for $E>E_c$ with $E_c=-1.54$ (first turning point of energy). The
critical
energy $E_c$ is the
counterpart of the Emden energy for box-confined
isothermal spheres. It corresponds to $k_{MCE}=7.44$. For $E<E_c$ there is no equilibrium state and the system
undergoes a gravothermal catastrophe (see the arrow in Fig. \ref{ebAhenon}). As
recalled in the Introduction, in the case of globular clusters,
core collapse leads to the formation
of a binary star surrounded by a hot halo.\footnote{In CE,
the system undergoes an isothermal collapse for $T<T_c$ with $T_c=0.613$
(first turning point of temperature)
corresponding to $k_{CE}=1.34$. In that
case, the isothermal collapse leads to a Dirac peak containing all the mass
\cite{sc2002,sc2004,ijmpb}.}

We now investigate the thermodynamical stability of the classical King model
according to the optimization problem (\ref{cf7}). The stable part
of the series of equilibria defines the microcanonical caloric curve.

For $E\rightarrow 0$, the system is stable in MCE
since it is equivalent to a polytrope of index $n=5/2$. Using the Poincar\'e theory
(see, {\it e.g.}, \cite{katzpoincare,ijmpb}), we
conclude that the series of
equilibria is stable until the first turning point of energy MCE and that it becomes
unstable after that point. In other words, the King distribution is an entropy maximum (EM)
at fixed mass and energy for $k<k_{MCE}$ and a saddle point (SP) of entropy
at fixed mass and energy for $k>k_{MCE}$. Since the series of equilibria always
rotates clockwise, a mode of stability is lost at each turning point of energy,
so the system is more and more unstable as $k$
increases.\footnote{There exist a region of ensemble inequivalence
between points CE and MCE in Fig. \ref{ebAhenon}, {\it i.e.} for configurations
with
$k_{CE}<k<k_{MCE}$ where $k_{CE}=1.34$ and $k_{MCE}=7.44$.
This part of the series of equilibria is stable in the microcanonical ensemble
(entropy maxima at fixed mass and energy) but unstable in the canonical
ensemble (saddle points of free energy at fixed mass). It corresponds to configurations with negative specific
heats $C=dE/dT<0$. We know that such configurations are forbidden in the
canonical
ensemble while they are allowed in the microcanonical ensemble. These results
are very similar to those obtained for box-confined isothermal spheres (see,
{\it e.g.}, \cite{paddy,katzrevue,ijmpb} for reviews).} Since there
is no global entropy maximum at fixed mass and energy for
classical self-gravitating systems (see footnote 3), the configurations
with $k<k_{MCE}$ are only metastable (local entropy maxima LEM). However, the probability to cross the
barrier of entropy and leave  a metastable state scales as $e^{-N}$ \cite{metastable,ijmpb}. For globular
clusters for which $N\sim 10^6$ this probability is totally negligible. Therefore, in practice, metastable
states are stable states. In this sense, globular clusters  can
be at statistical equilibrium, described
by the King distribution with $k<k_{MCE}$, even if there is no equilibrium state
in a strict
sense.  Their lifetime
is controlled by evaporation and core collapse as recalled in the
Introduction \cite{bt}. 

In Fig. \ref{esAhenon} we plot the entropy $S$ normalized by $M$ as a function of the energy $-E$. Since $\delta S=\beta\delta E$ (for a fixed mass $M$) in MCE, we find that
$S(k)$ is extremum when $E(k)$ is extremum. This explain the ``spikes'' observed
in Fig. \ref{esAhenon}. The series of equilibria becomes unstable after the first
spike. This is in agreement with the fact that the states on the unstable
branches (after the first spike) have lower entropy than the states on the
stable branch (before the first spike) for the same energy.

\section{Comparison with the observations of dark matter halos}
\label{sec_comp}

The fermionic King model provides a realistic model of dark matter halos made of
massive neutrinos. Large dark matter halos are non-degenerate and the classical
King model can be used. Figs. \ref{densityLOG}-\ref{vitesseLIN} compare the
prediction of the King model with the empirical Burkert profile \cite{burkert}:
\begin{equation}
\frac{\rho(r)}{\rho_0}=\frac{1}{(1+x)(1+x^2)},\qquad x=\frac{r}{r_h},
\label{dm1}
\end{equation}
that fits a large variety of dark matter halos. We see that the Burkert profile
is relatively close to the King profile at, or close to, the limit of
microcanonical stability $k_{MCE}=7.44$. This agreement may be understood as
follows. Because of collisions and evaporation, the concentration parameter
$k(t)$ increases with time while the slope $\alpha(t)$ of the halo profile
decreases until an instability takes place \cite{cohn}. Therefore, we expect
that the halos
that have not collapsed have a value of $k$ close to its maximum stable value
$k_{MCE}$, corresponding to a slope $\alpha\sim 3$. At that point, the King
profile can be fitted by the modified Hubble profile \cite{bt}:
\begin{equation}
\frac{\rho(r)}{\rho_0}=\frac{1}{\lbrack 1+(4^{2/3}-1) x^2\rbrack^{3/2}},\qquad
x=\frac{r}{r_h}.
\label{mh4}
\end{equation}
It has a flat core and a halo in which the density decreases as $r^{-3}$. These
properties are qualitatively similar to the properties of the Burkert profile.

Therefore, the King model gives a relevant description of dark matter halos without fitting parameter. Furthermore, the marginal King profile is physically justified contrary to the Burkert profile that is purely empirical. If we compare the Burkert profile and the marginal King profile more precisely, we see that the agreement is very good in the core for $r\le r_h$.
Therefore, it appears that the core of dark matter halos
is isothermal.\footnote{We must be careful, however,
because many
models of dark matter halos give a good agreement with the Burkert profile for
$r<r_h$.} We note
that the classical King model, which is a truncated isothermal sphere, produces
a flat density profile in the core, instead of a cusp, without the need to
invoke quantum mechanics. Therefore, warm dark matter (WDM) may account for the
absence of density cusps in observations. This thermalisation may be due to
collisionless violent relaxation \cite{lb,csmnras}, not to collisional relaxation, as explained in the Introduction.

The agreement with the observed rotation
curves of galaxies is less good in the halo for $r\ge r_h$. This discrepancy
may be interpreted
as a result of an incomplete relaxation, as in the case of stellar systems
\cite{lb}. Therefore, the cores of dark matter halos seem to be isothermal but
deviations appear in the halo.
The same observation is made for elliptical galaxies. As a whole, these considerations show
that statistical mechanics provides a good starting point to understand the structure
of dark matter halos but that more work remains to be done in order to obtain a complete picture.

\section{Thermodynamics of the fermionic King model}
\label{sec_proper}

We now consider the thermodynamics of the fermionic King model for arbitrary
values of the degeneracy parameter $\mu$. This distribution function is
appropriate to describe dark matter halos of various sizes if they are made of massive neutrinos or have experienced violent relaxation.

\begin{figure}
\begin{center}
\includegraphics[clip,scale=0.3]{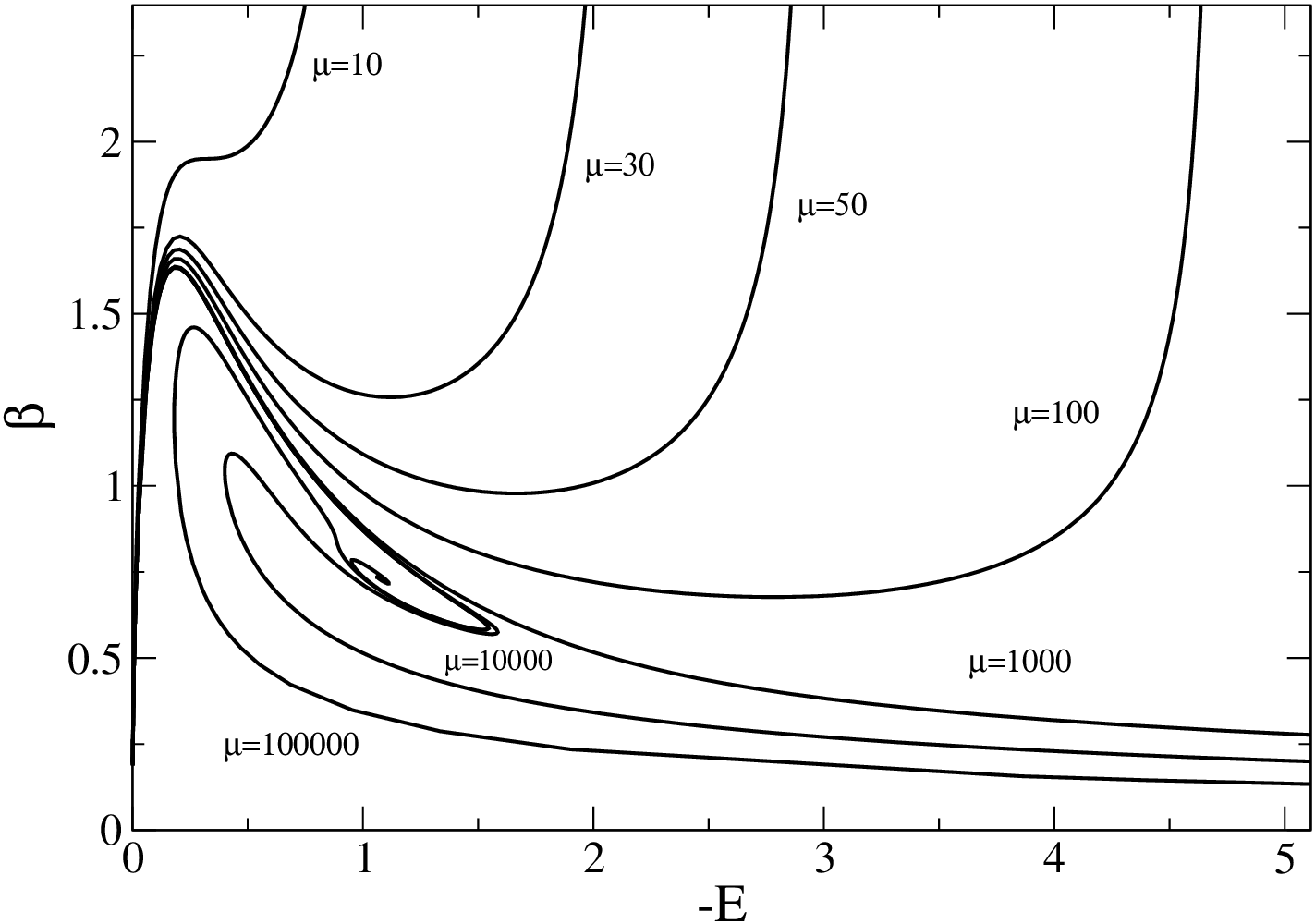}
\caption{Series of equilibria corresponding to the fermionic King model for different values of the degeneracy parameter $\mu$ (note that for
large values of $\mu$, the minimum energy $E_{min}(\mu)$ corresponding to
$T=0$ is outside the frame of the Figure). For $\mu\gg 1$,
the series of equilibria makes several rotations before unwinding. }
\label{calomulti}
\end{center}
\end{figure}

When an exclusion constraint is taken into account (in the sense of Pauli or
in the
sense of Lynden-Bell), the structure of the
series of equilibria $\beta(E)$ depends on the value of the degeneracy
parameter $\mu$ as shown in Fig. \ref{calomulti}. The degeneracy parameter $\mu$
is a measure of the size of the system \cite{cml}. Large values of $\mu$
correspond to large halos and small values of $\mu$ correspond to small halos. 
For $\mu\rightarrow
+\infty$ we recover the classical spiral of
Fig. \ref{ebAhenon}. However, for smaller values of $\mu$, we see
that the effect of the exclusion constraint  is to unwind the spiral.  Depending on
the value of the degeneracy parameter, the series of equilibria can
have different shapes.

\begin{figure}
\begin{center}
\includegraphics[clip,scale=0.3]{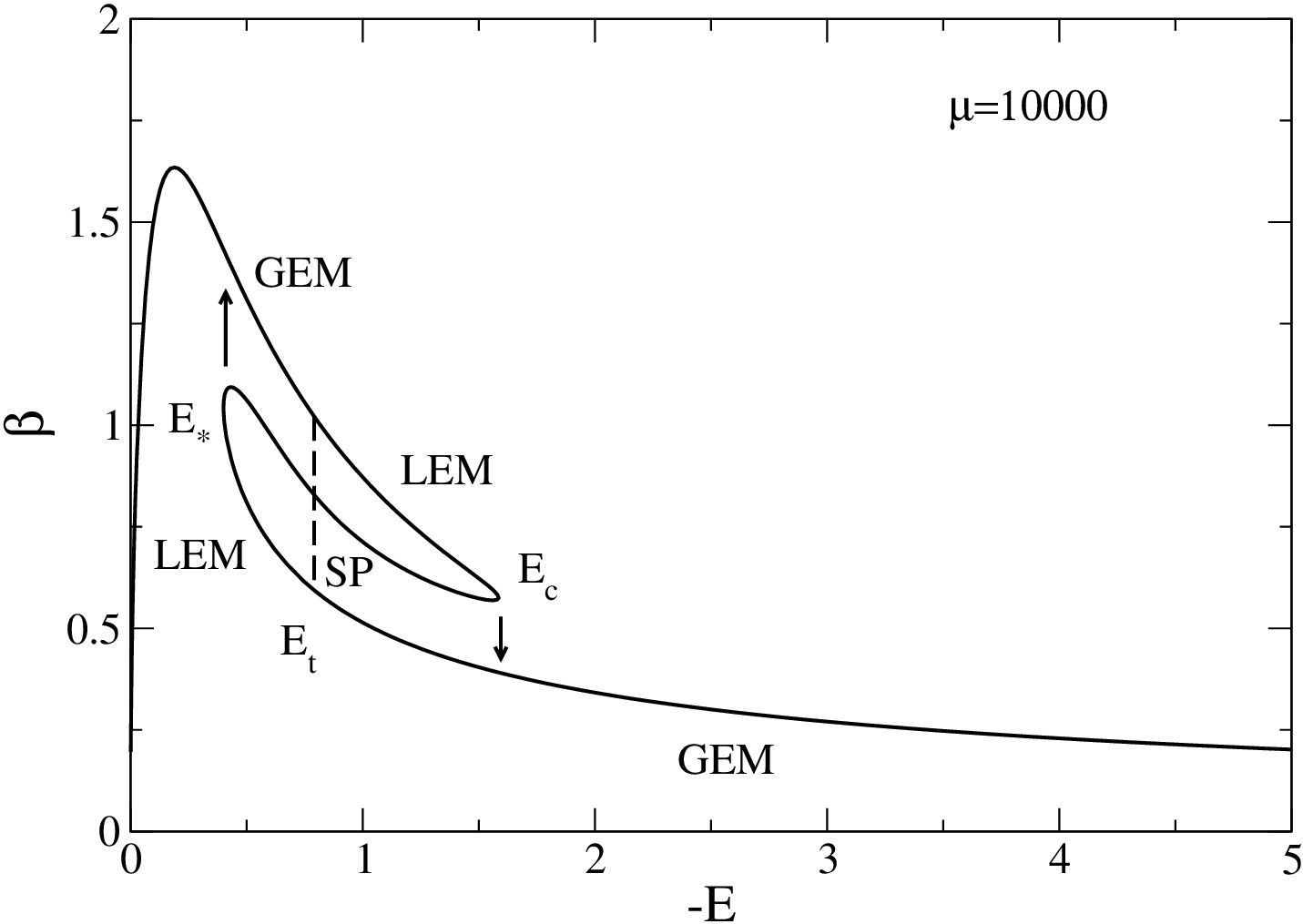}
\caption{Series of equilibria corresponding to the fermionic King model with $\mu=10000$.}
\label{ETmicro5summary}
\end{center}
\end{figure}

For $\mu=10000$ (large halos), the series of equilibria is represented in Fig.
\ref{ETmicro5summary}. It has a $Z$-shape structure.
Since the curve $\beta(E)$ is multi-valued, this gives rise to microcanonical phase
transitions. For $k\rightarrow 0$, the series of equilibria is stable.
According to the Poincar\'e theory \cite{katzpoincare,ijmpb}, it remains
stable until the first turning point of
energy. At that point a mode of stability is lost as the series of equilibria rotates
clockwise. However, the stability is re-gained at the second turning point of
energy since the series of equilibria rotates anti-clockwise \cite{ijmpb}.

\begin{figure}
\begin{center}
\includegraphics[clip,scale=0.3]{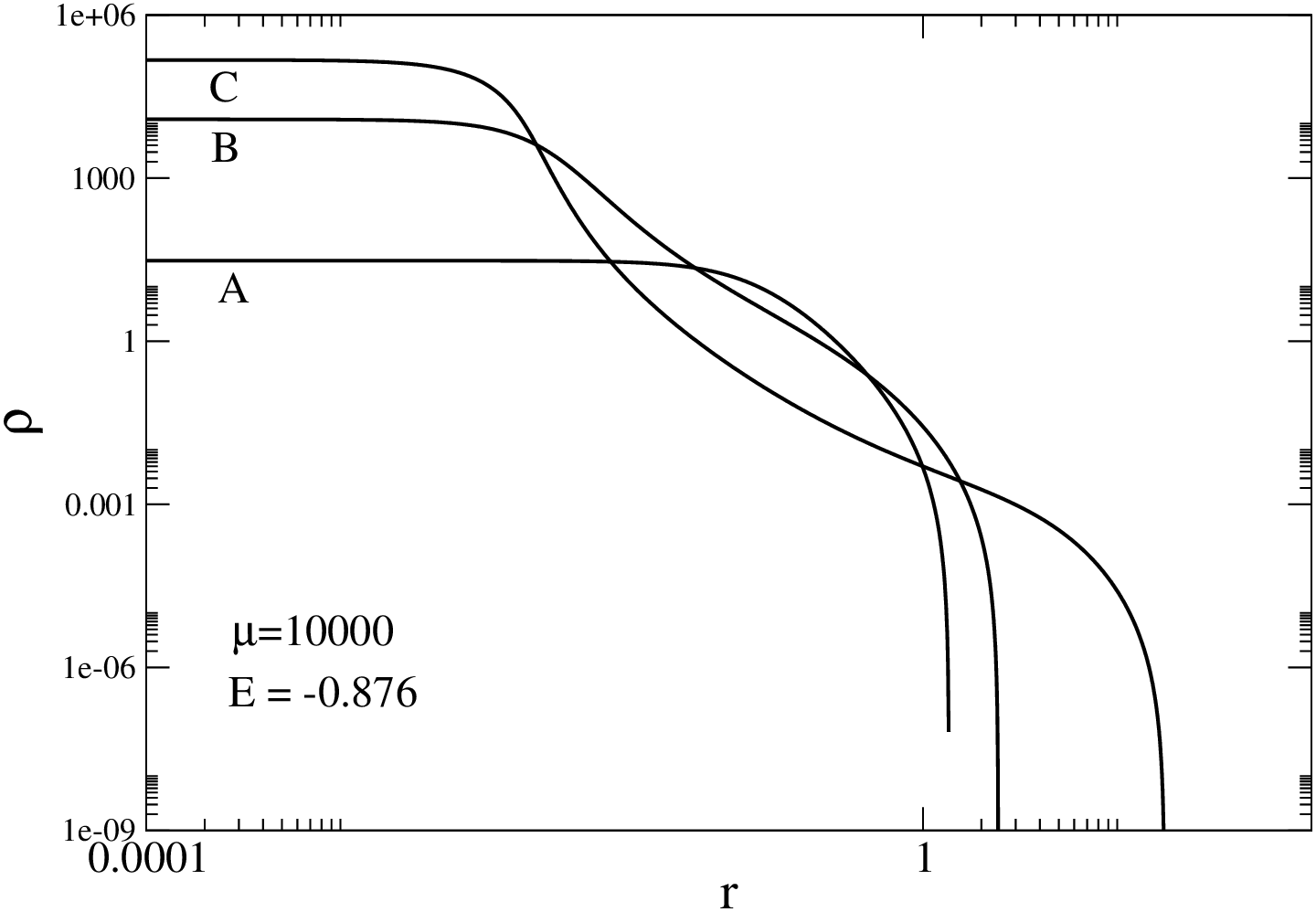}
\caption{Normalized density profiles corresponding to the different phases of the fermionic
King model with $\mu=10000$ and $E=-0.876$. The radial distance is scaled by $1/(4\pi G M^{1/3}A^{2/3})$
and the density by $(4\pi G)^3A^2M^2$. The core-halo structure of solutions B
and C comprising a dense
degenerate nucleus surrounded by an atmosphere is clearly visible.}
\label{profilesREAL}
\end{center}
\end{figure}

\begin{figure}
\begin{center}
\includegraphics[clip,scale=0.3]{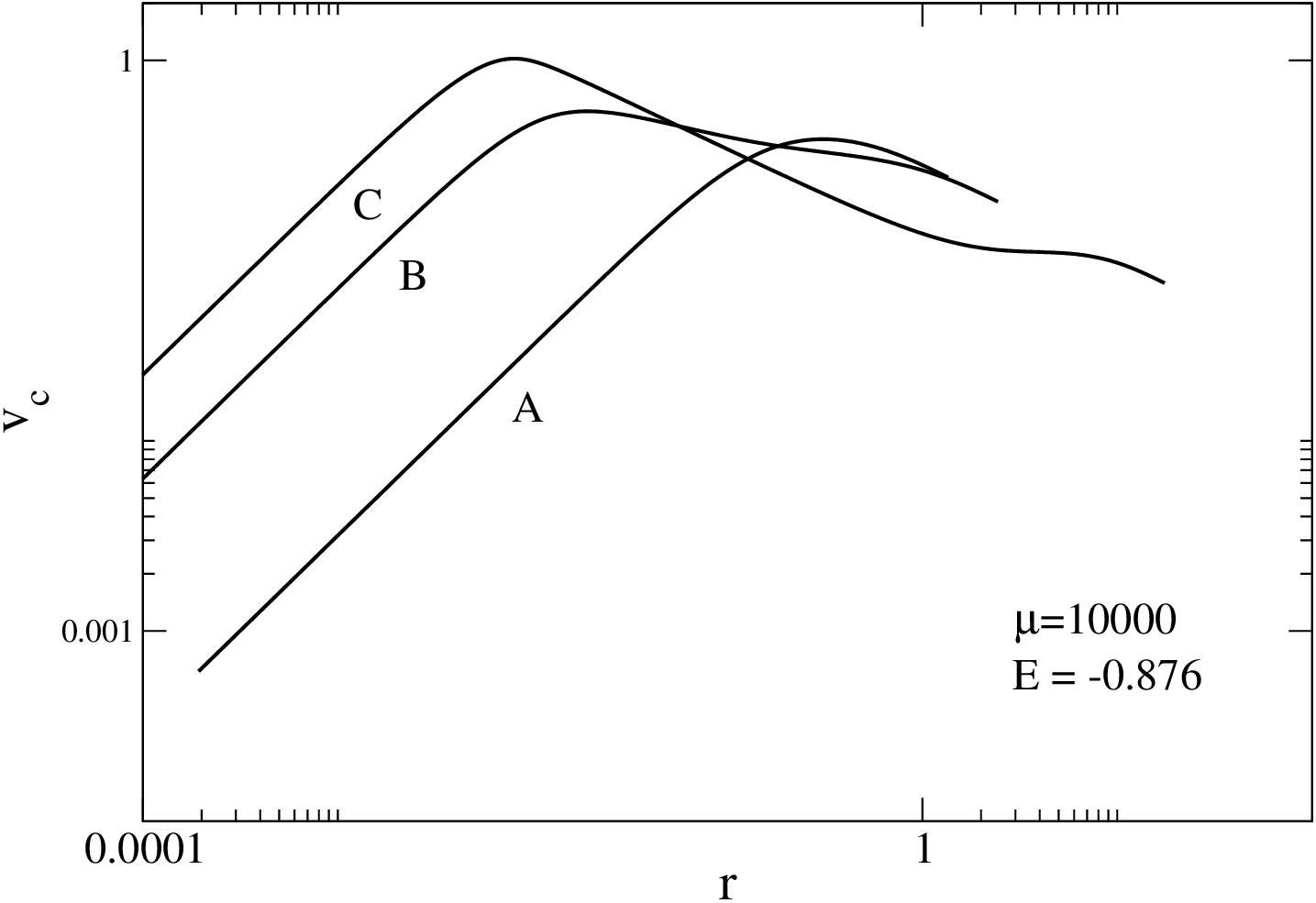}
\caption{Normalized rotation curves corresponding to the different phases of the fermionic
King model with $\mu=10000$ and $E=-0.876$. The radial distance is scaled by $1/(4\pi G M^{1/3}A^{2/3})$
and the circular velocity by $4\pi GA^{1/3}M^{2/3}$.}
\label{profilesVREAL}
\end{center}
\end{figure}

The solutions on the upper branch are stable (entropy maxima EM). They  are non degenerate
and have a smooth density profile. They form the ``gaseous phase'' (see solution $A$
in Figs. \ref{profilesREAL} and \ref{profilesVREAL}).
The solutions on the lower branch are also stable (entropy maxima EM). They  have a core-halo structure
consisting of  a degenerate nucleus surrounded by a dilute  ``atmosphere''.  They
form the ``condensed phase'' (see solution $C$ in Figs. \ref{profilesREAL} and \ref{profilesVREAL}).
The solutions on the intermediate  branch are unstable (saddle points of entropy SP). They
are similar to the solutions of the gaseous phase but they contain a small embryonic
degenerate nucleus playing the role of a ``germ'' in the theory of phase transitions (see solution $B$ in
Figs. \ref{profilesREAL} and \ref{profilesVREAL}). These solutions form a
barrier of entropy that the
system has to cross in order to pass from the gaseous phase to the condensed phase,
or inversely (see Ref. \cite{ijmpb} for more details).

\begin{figure}
\begin{center}
\includegraphics[clip,scale=0.3]{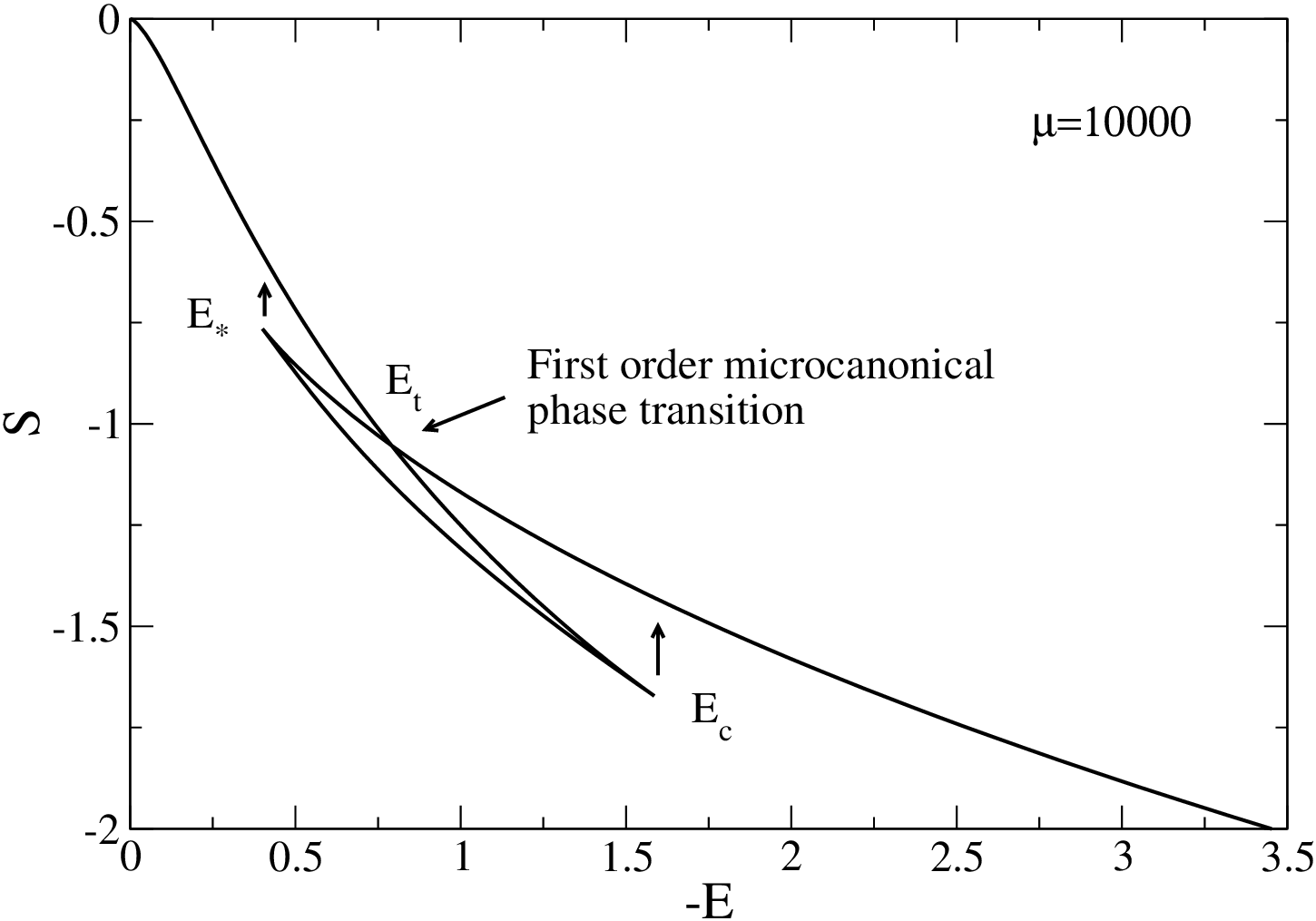}
\caption{Entropy of each phase versus
energy for $\mu=10000$.}
\label{Smicro5}
\end{center}
\end{figure}

If we compare the entropy of the solutions (see Fig. \ref{Smicro5}), we expect a first
order microcanonical phase transition to take place at a transition energy
$E_t(\mu)$ where the entropy of the gaseous phase and the entropy of the
condensed phase become equal. It is marked by a discontinuity of the inverse
temperature $\beta=\partial S/\partial E$ (first derivative of the entropy) in the strict caloric curve.
The transition energy $E_t(\mu)$ may also be obtained by performing a Maxwell  construction
(see the vertical plateau in Fig. \ref{ETmicro5summary}) \cite{ijmpb}. For $E>E_t$, the gaseous phase is fully stable (global entropy maximum GEM at fixed
mass and energy) while the condensed phase is metastable (local entropy maximum LEM
at fixed mass and energy). For $E<E_t$, the gaseous phase is metastable (LEM) while
the condensed phase is fully stable (GEM). However, for systems with long-range
interactions, the metastable states have
considerably large lifetimes, scaling as $e^N$, so that the first order
microcanonical phase transition does not take place in practice \cite{metastable,ijmpb}. Therefore, the physical caloric curve must take metastable states into account.

\begin{figure}
\begin{center}
\includegraphics[clip,scale=0.3]{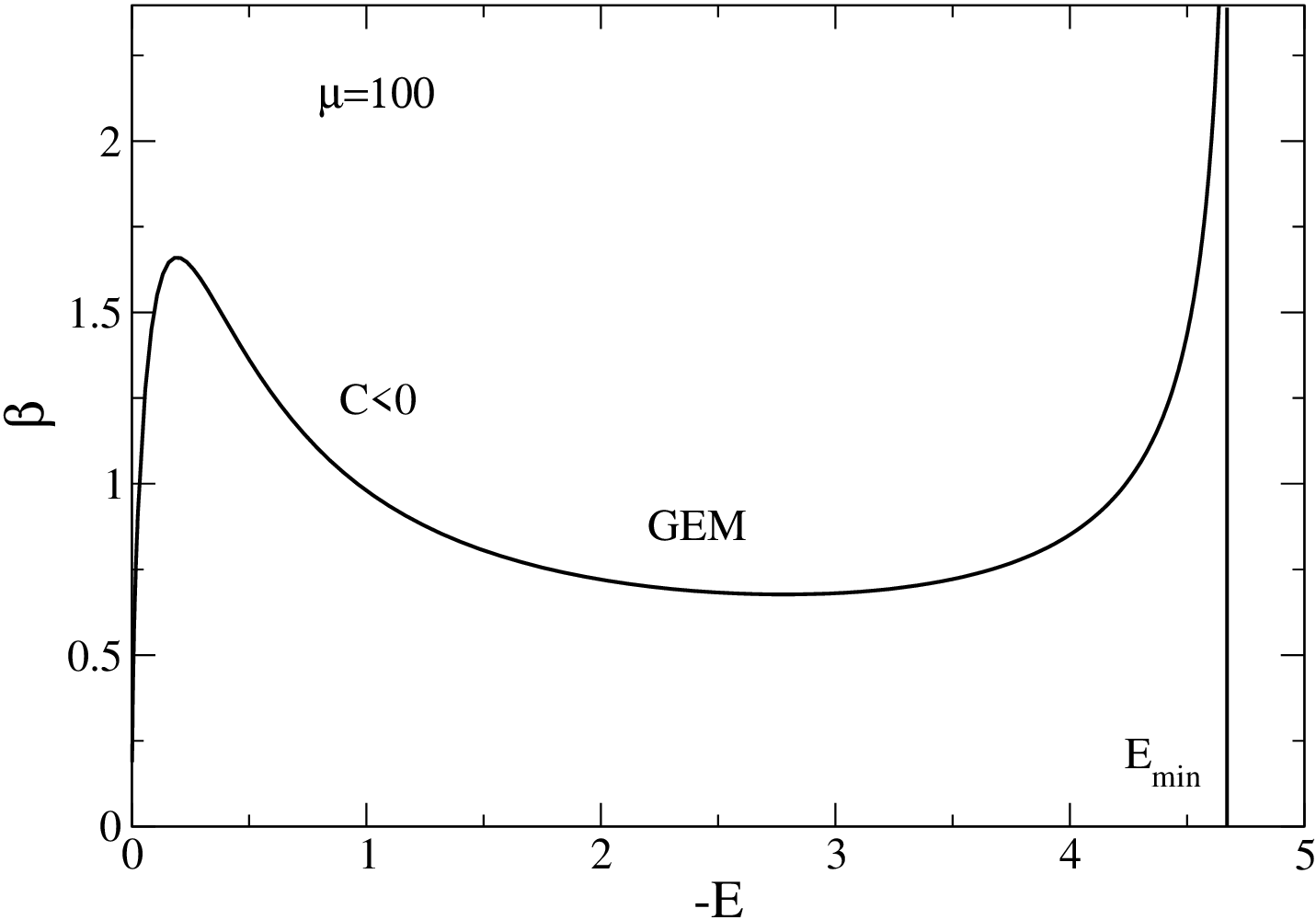}
\caption{Series of equilibria corresponding to the fermionic King model with $\mu=100$.}
\label{ETmicro3}
\end{center}
\end{figure}

\begin{figure}
\begin{center}
\includegraphics[clip,scale=0.3]{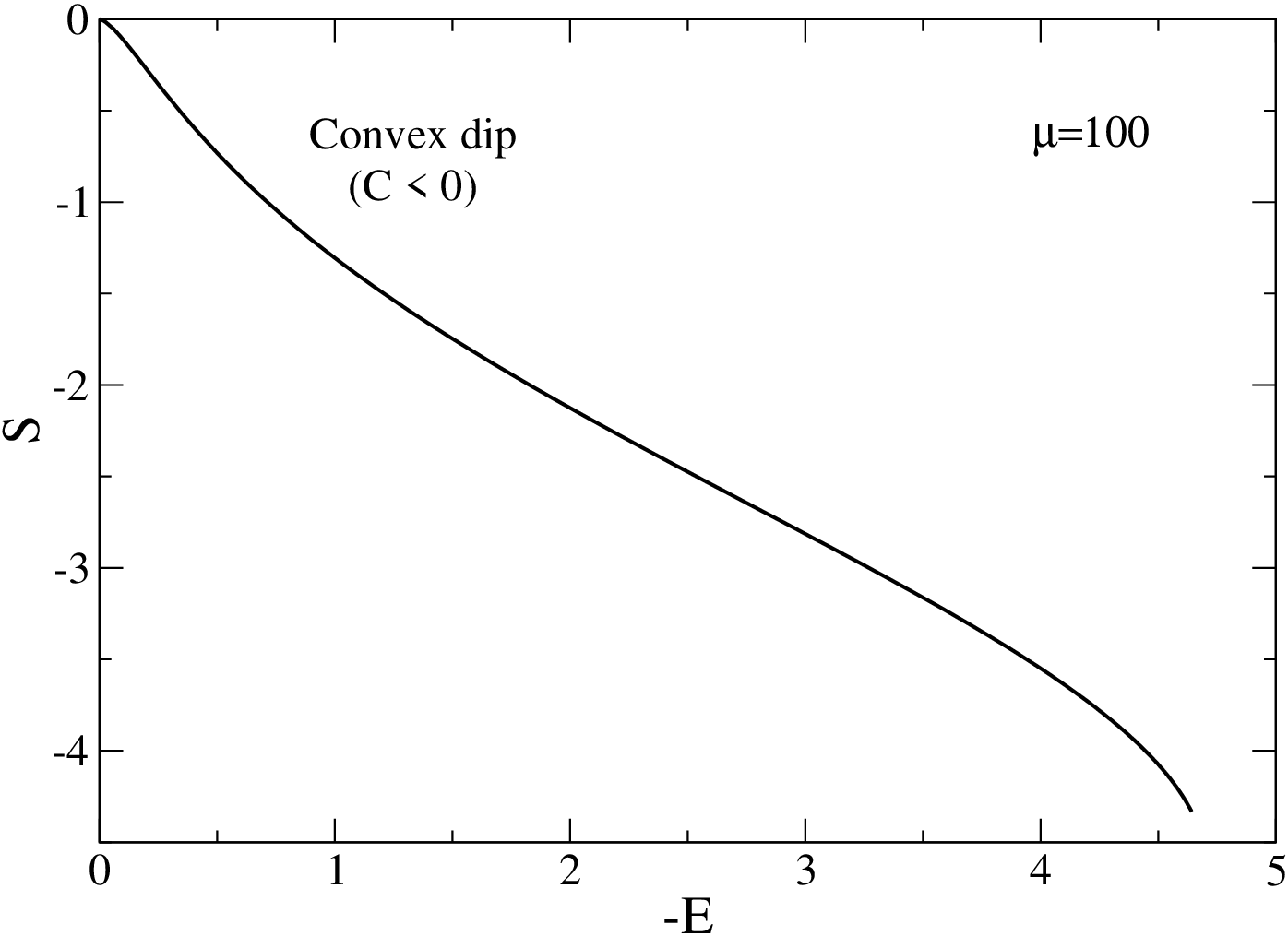}
\caption{Entropy versus
energy for $\mu=100$.}
\label{Smicro3}
\end{center}
\end{figure}

\begin{figure}
\begin{center}
\includegraphics[clip,scale=0.3]{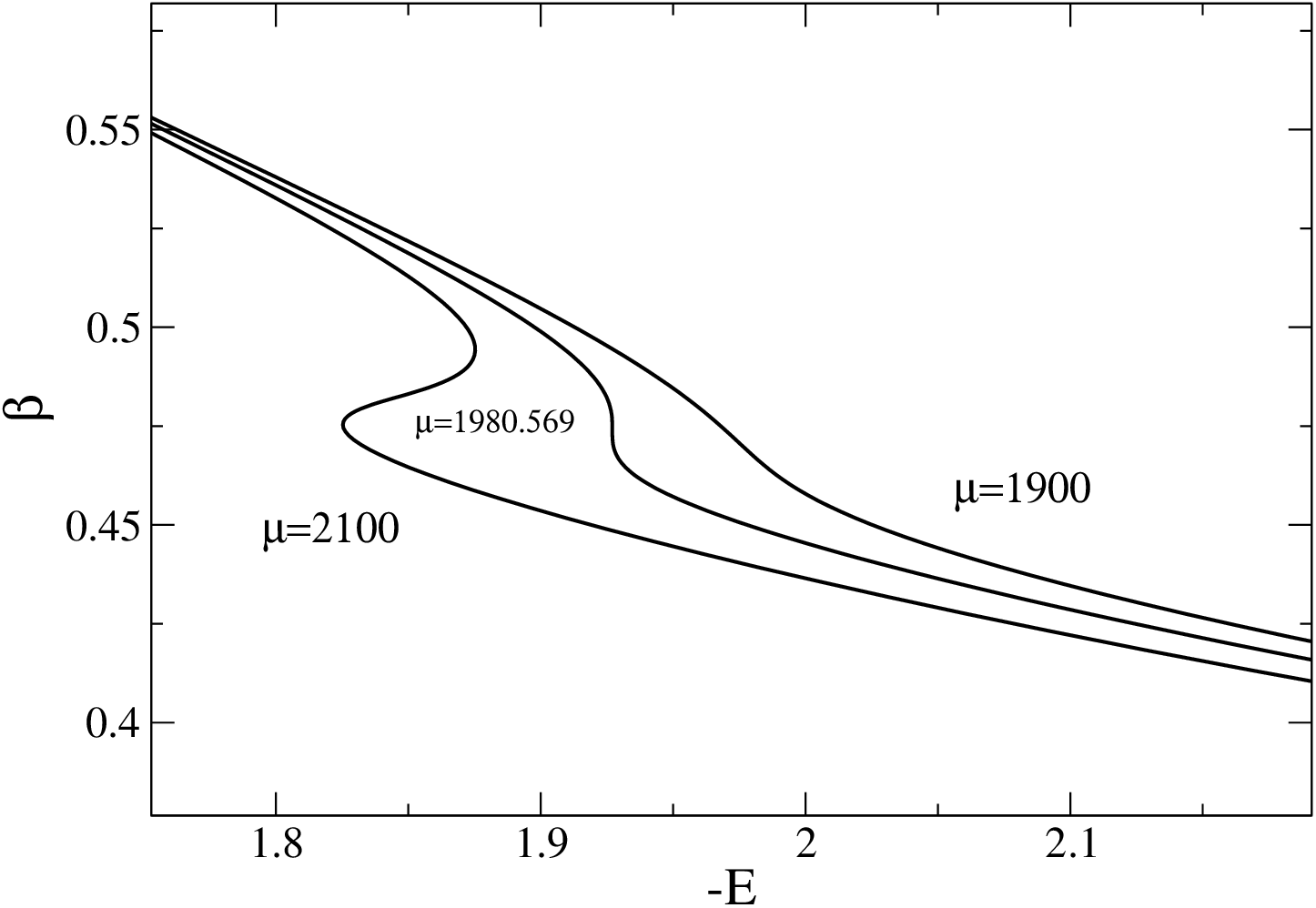}
\caption{Series of equilibria close to the microcanonical critical point
$\mu_{MCP}=1980$ below which microcanonical
phase transitions disappear.}
\label{mcp}
\end{center}
\end{figure}

At $E=0^{-}$, the system is in the gaseous phase. If we decrease the
energy it remains in the gaseous phase until the critical energy $E_c(\mu)$ at
which the gaseous phase disappears (spinodal point).  For sufficiently large
values of $\mu$, this is close to  the critical energy $E_{c}=-1.54$ obtained
with the classical King model. For $E<E_c(\mu)$,
the system undergoes gravitational collapse (gravothermal catastrophe). However, the
collapse stops when the core of the system becomes degenerate (see Fig. \ref{ETmicro5summary}).
In that case, it
ends up in the condensed phase. The system has a core-halo structure with a
degenerate nucleus surrounded by a non-degenerate atmosphere. Since the
collapse is
accompanied by a discontinuous jump of entropy (see Fig. \ref{Smicro5}),
this is sometimes called a microcanonical zeroth order phase
transition. If we now increase the energy, the system remains in the condensed
phase until the critical energy $E_*(\mu)$ at which the condensed phase
disappears. For $E>E_*(\mu)$, the system undergoes an ``explosion'' reversed to
the collapse
and returns to the gaseous phase (see Fig. \ref{ETmicro5summary}). In this sense, we have described an hysteretic
cycle in the microcanonical ensemble (see the arrows in Figs. \ref{ETmicro5summary} and \ref{Smicro5}).

For $\mu=100$ (small halos), the series of equilibria is represented in Fig.
\ref{ETmicro3}. It has an $N$-shape structure. Since the  curve $\beta(E)$ is
univalued there is no phase transition in MCE. All the solutions are
fully stable (global entropy maxima GEM at fixed mass and energy), up to the
minimum energy $E_{min}$ at which $\beta\rightarrow +\infty$. For intermediate
energies,
the caloric curve displays a region of negative specific heats
($C=dE/dT<0$).\footnote{In CE,
this region of negative specific heats is replaced by a canonical phase transition
connecting the gaseous phase (left branch) to the condensed phase (right branch)
\cite{ijmpb,cml}.}

Another interesting curve is the entropy versus energy relation $S(E)$
represented in Fig. \ref{Smicro3}. It displays  a convex
intruder in the region of negative specific heats.

Microcanonical phase transitions occur for $\mu>\mu_{MCP}=1980$ (microcanonical
critical point). They are associated with the turning point
of energy at $E_c$ and the multi-valuedness of the series of equilibria
$\beta(E)$. For $\mu<\mu_{MCP}$,
the series of equilibria $\beta(E)$ unwinds and becomes univalued (see Fig. \ref{mcp}). In that case, there is no microcanonical
phase transition (no gravothermal catastrophe) anymore.

%\section{Can large dark matter halos harbor a fermion ball?}
%\label{sec_canwe}

\section{Density profiles and rotation curves of the fermionic King model}

In this section, we discuss in greater detail the structure of the density profiles and of the
rotation curves of the fermionic King model.

\begin{figure}
\begin{center}
\includegraphics[clip,scale=0.3]{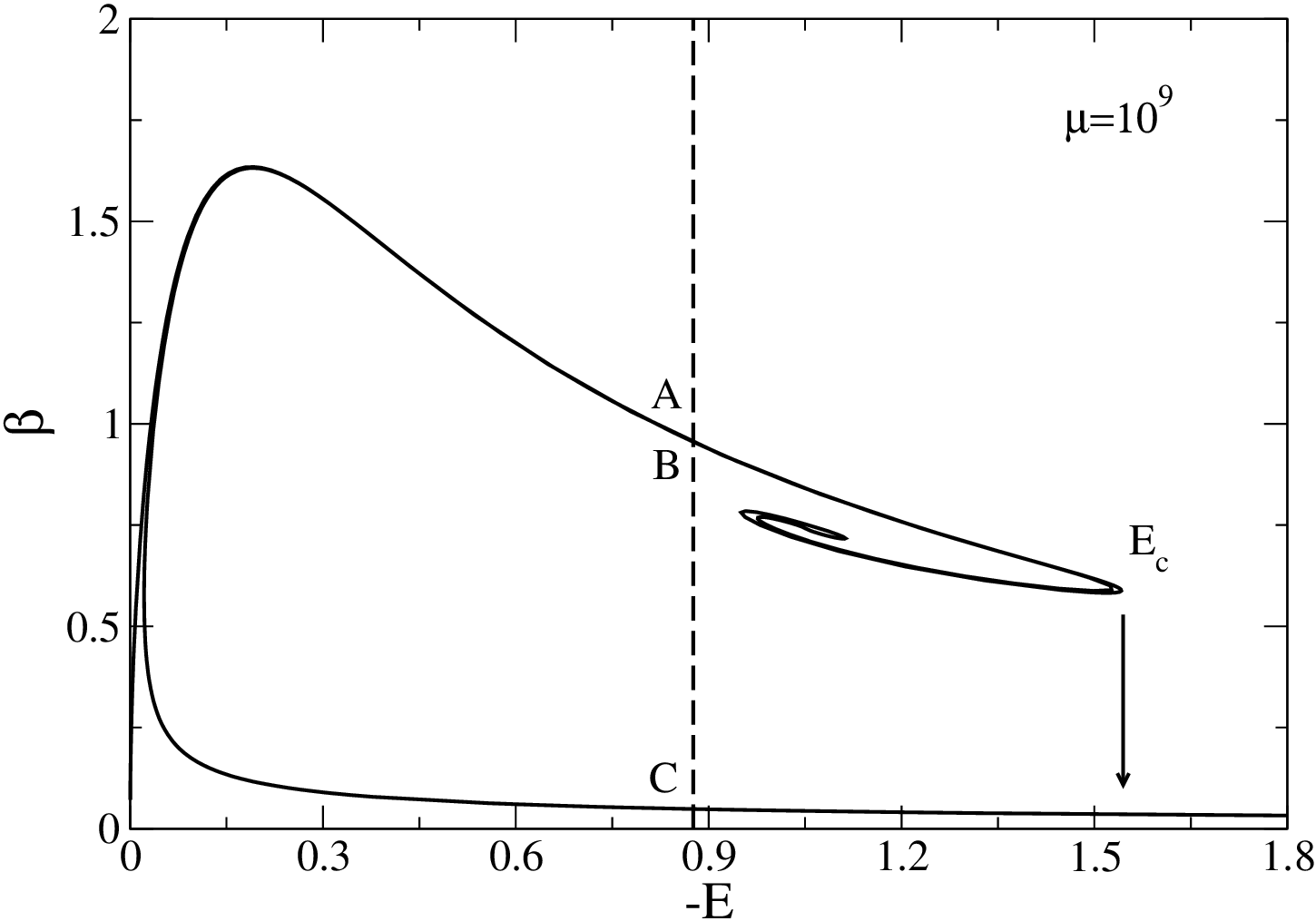}
\caption{Series of equilibria corresponding to the fermionic King model with $\mu=10^9$.}
\label{mutresgrandprolongehenon}
\end{center}
\end{figure}

We first take an energy $E=-0.876>E_c$ and study the evolution of the
solutions A, B and C as $\mu$ increases. The series of equilibria is represented in
Fig. \ref{mutresgrandprolongehenon} for $\mu=10^9$.
For $\mu\rightarrow +\infty$, the branches A and B superimpose
while the branch C coincides with the $\beta=0$ axis. As
a result, we recover the spiral of Fig. \ref{ebAhenon}.

\begin{figure}
\begin{center}
\includegraphics[clip,scale=0.3]{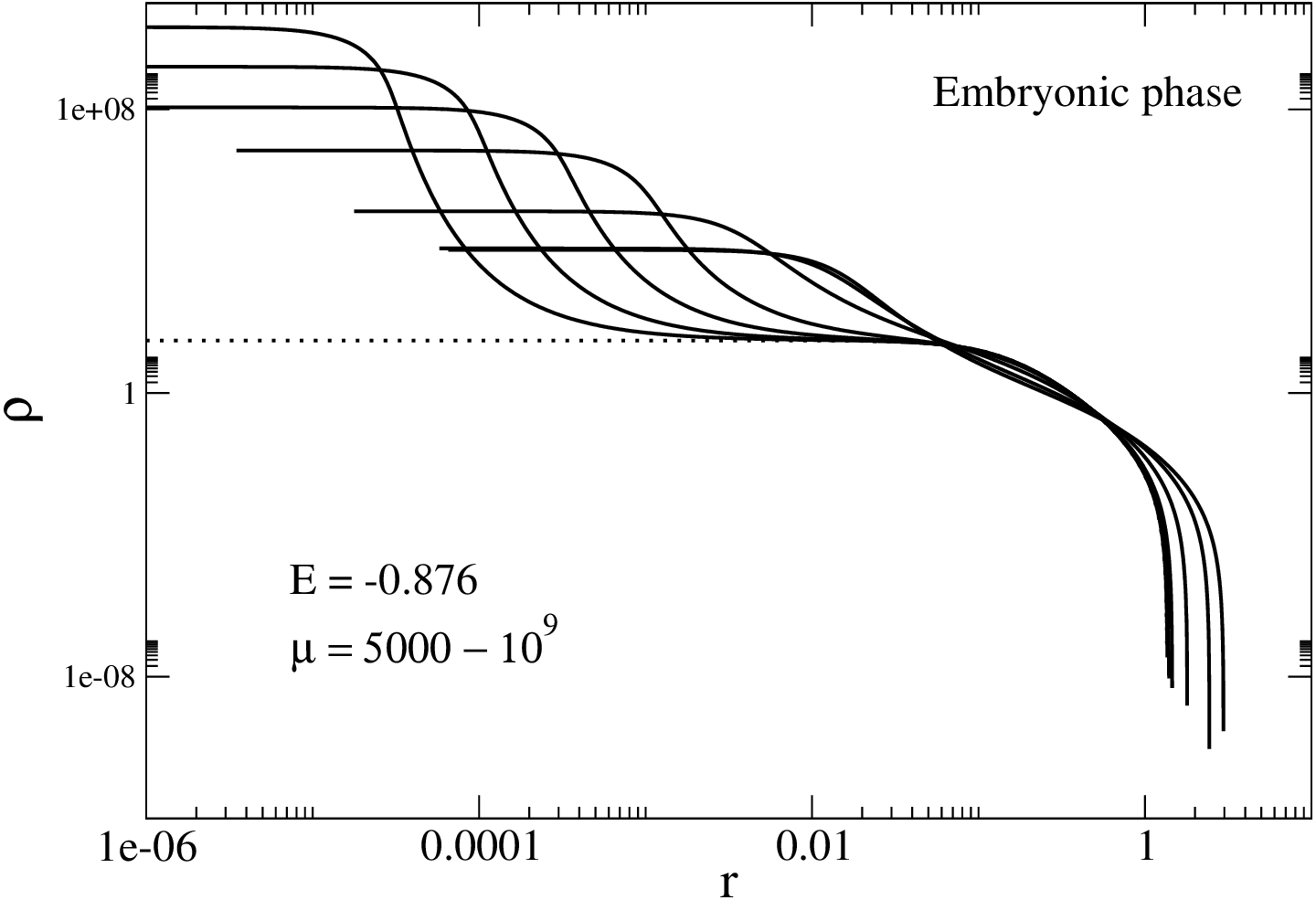}
\caption{Density profile of the embryonic phase (solution B) for different
values of $\mu$ in logarithmic scales (we have selected $\mu=5000$, $10^4$,
$10^5$, $10^6$, $10^7$, $10^8$, and $10^9$).
For increasing $\mu$, the solution B coincides with the solution A (gaseous phase; dotted line)
corresponding to the classical King model, except that it contains a small
embryonic degenerate nucleus with
a small mass and a small absolute value of potential energy. This nucleus of almost constant density is followed by a plateau as detailed in \cite{csmnras}.}
\label{embryonRHOdot}
\end{center}
\end{figure}

\begin{figure}
\begin{center}
\includegraphics[clip,scale=0.3]{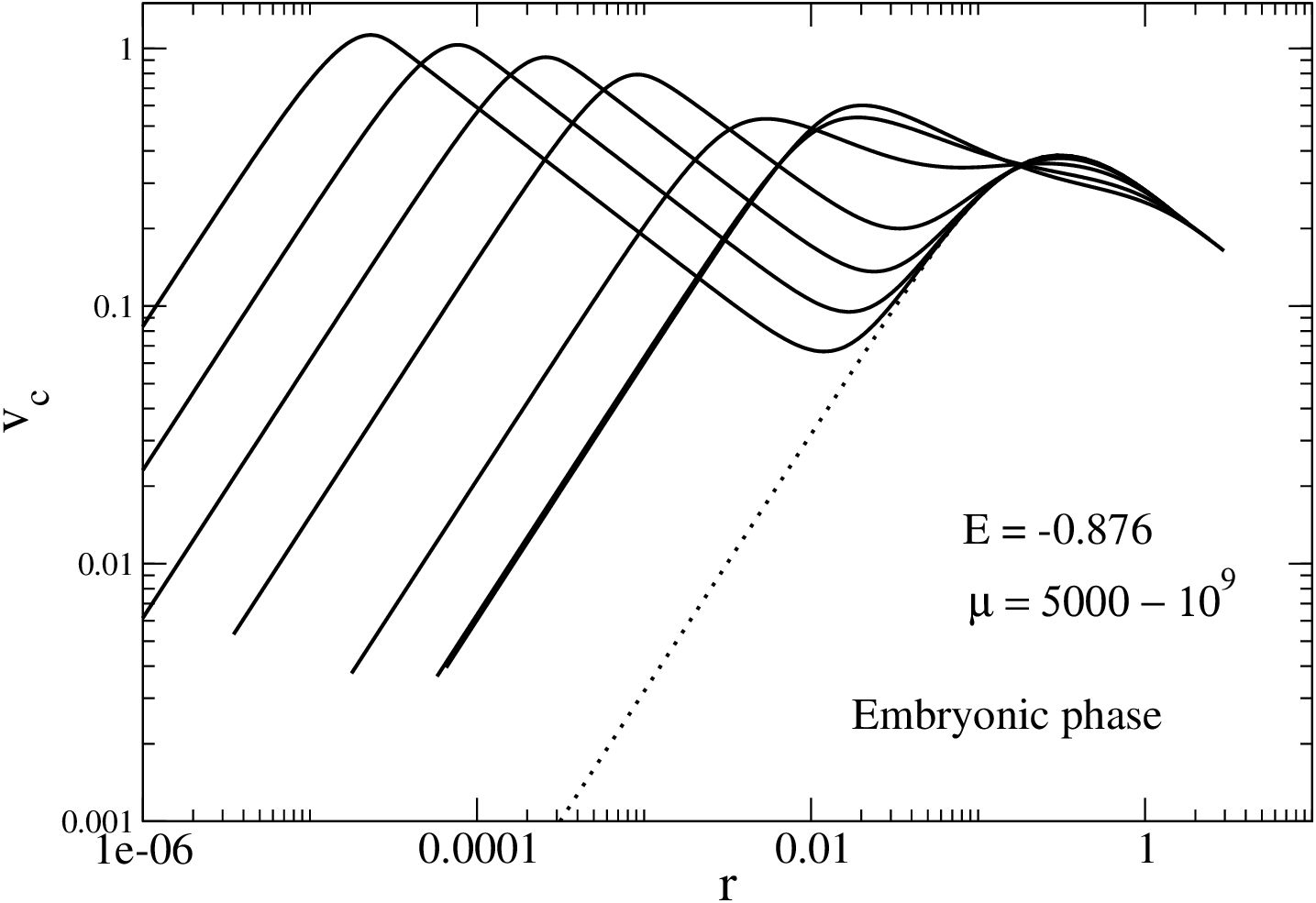}
\caption{Circular velocity of the embryonic phase (solution B) for different
values of $\mu$ in logarithmic scales.
For increasing $\mu$, the solution B approaches the solution A (gaseous phase; dotted line)
corresponding to the classical King model, except at very small distances. The presence of a small nucleus (fermion
ball)  where $v_c\propto r$ followed by a plateau where $v_c\propto r^{-1/2}$ manifests itself by a secondary peak in the rotation curve at the very
center
of the system (see Fig. \ref{embryonVlindot}). However,
these distances are probably not accessible to observations. Furthermore, these solutions are thermodynamically
unstable so this secondary peak may not be physical.}
\label{embryonVdot}
\end{center}
\end{figure}

\begin{figure}
\begin{center}
\includegraphics[clip,scale=0.3]{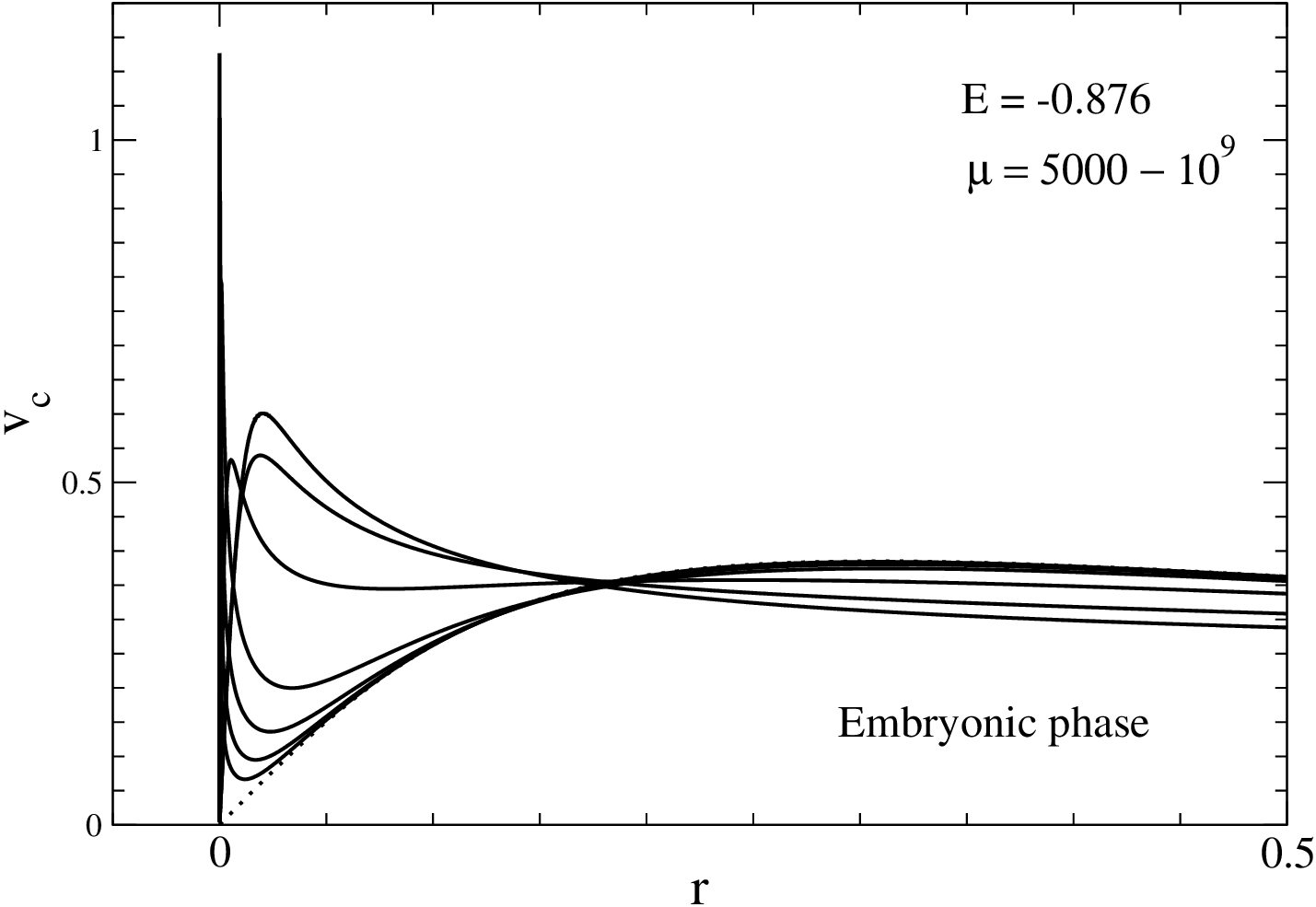}
\caption{Circular velocity of the embryonic phase (solution B) for different
values of $\mu$ in linear scales. For large values of
$\mu$, we recover the classical King model (dotted line) except at the very center. The secondary peak due to the
degenerate nucleus (fermion ball) manifests itself by a spike near the origin. }
\label{embryonVlindot}
\end{center}
\end{figure}

\begin{figure}
\begin{center}
\includegraphics[clip,scale=0.3]{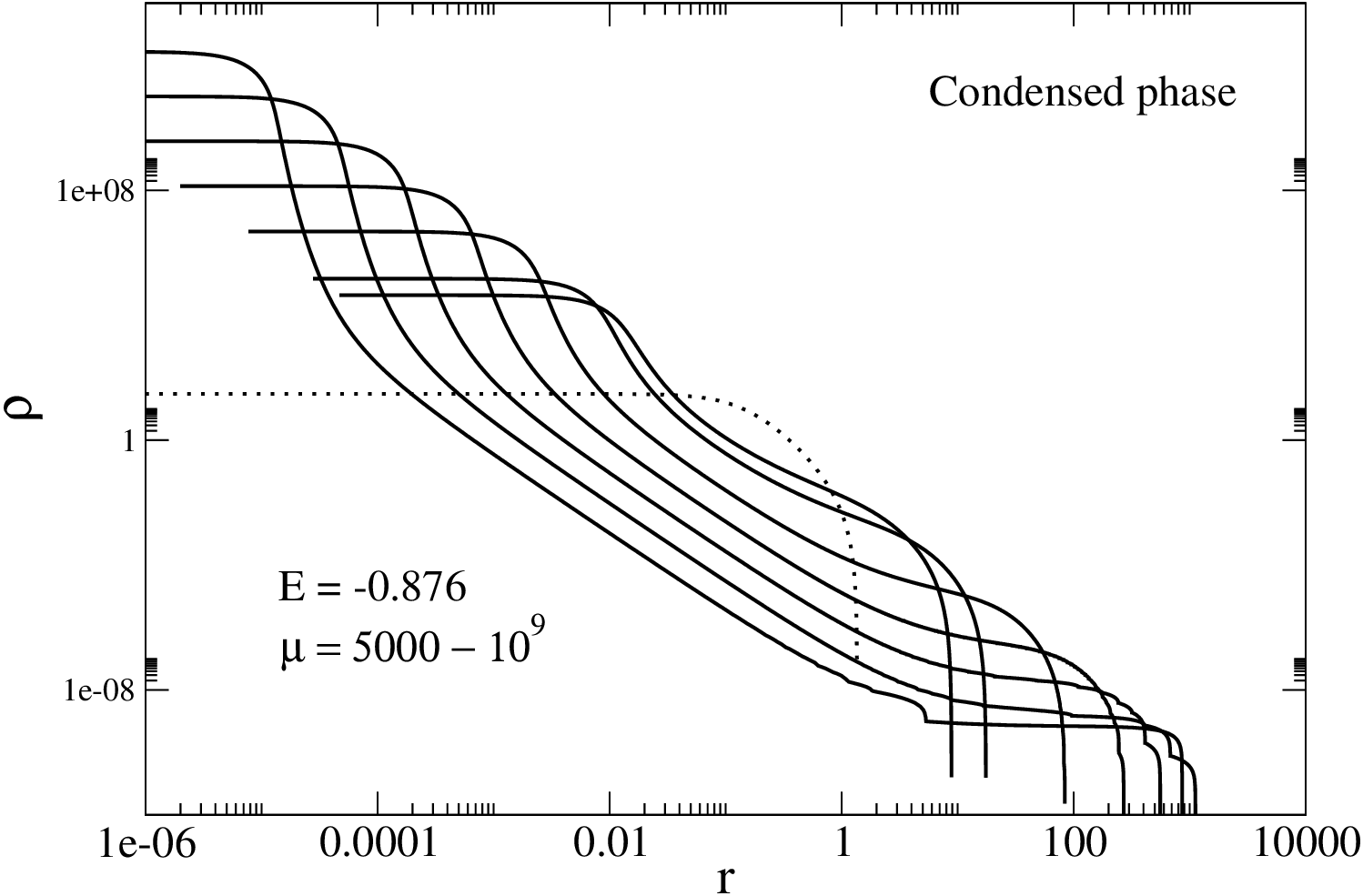}
\caption{Density profile of the condensed phase (solution C) for different
values of $\mu$ in logarithmic scales.
For increasing values of $\mu$,
the solution C contains a small degenerate nucleus with a relatively small mass
but a more and more negative
potential energy.  As a result, the halo becomes hotter and hotter in order to conserve the total energy. This is why it forms a sort of plateau with constant density that extends at larger and larger distances. The resulting profile is very
different from solution A (gaseous phase;
dotted line) corresponding to the classical King model.}
\label{condensedRHOdot}
\end{center}
\end{figure}

\begin{figure}
\begin{center}
\includegraphics[clip,scale=0.3]{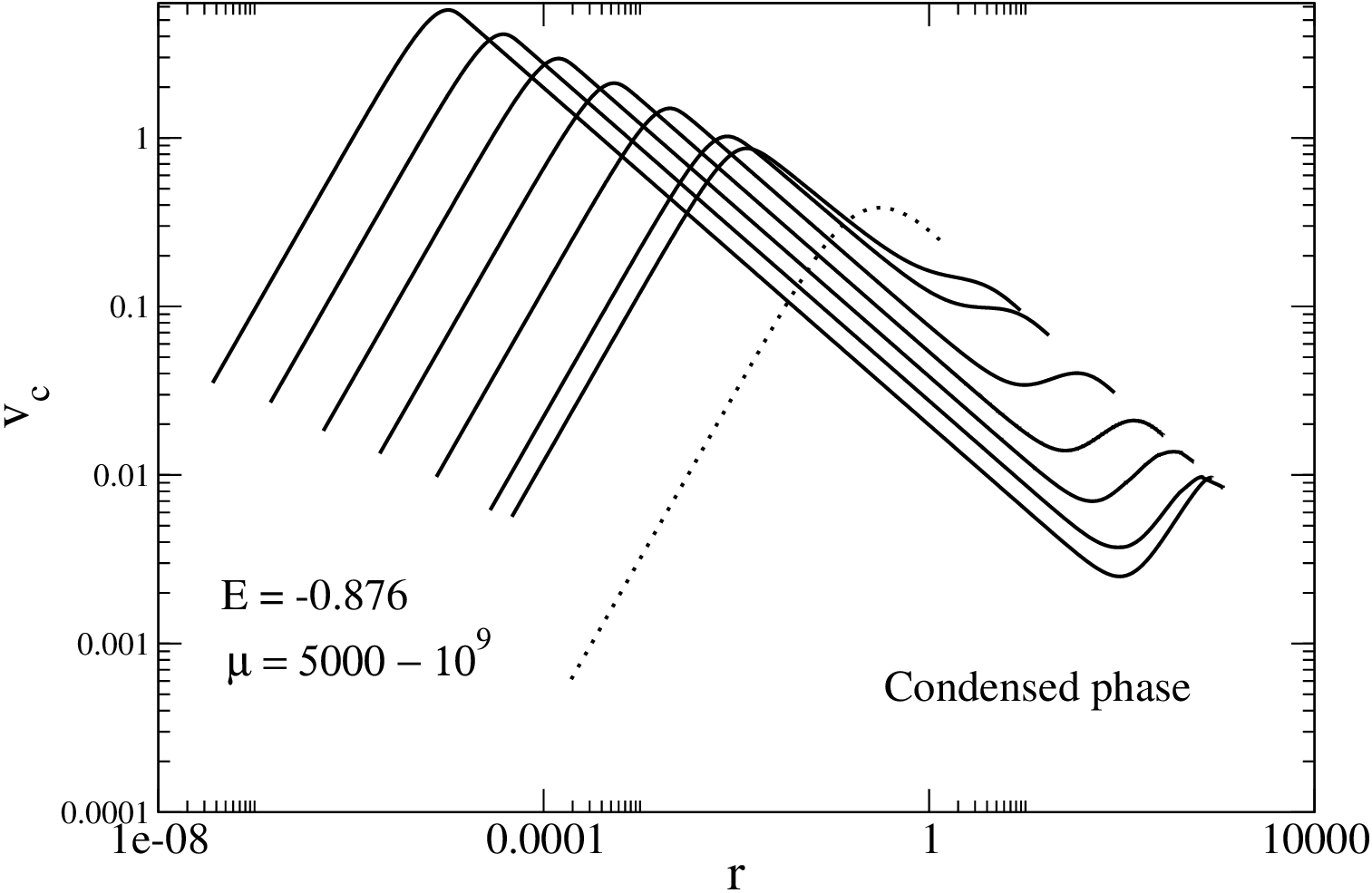}
\caption{Circular velocity of the condensed phase  (solution C) for different values of $\mu$ in logarithmic
scales.  It is very
different from solution A (gaseous phase; dotted line)  corresponding to the classical King model. This is
because the halo is expelled at large distances as the nucleus becomes denser
and denser,  and more and more energetic.}
\label{condensedVdot}
\end{center}
\end{figure}

\begin{figure}
\begin{center}
\includegraphics[clip,scale=0.3]{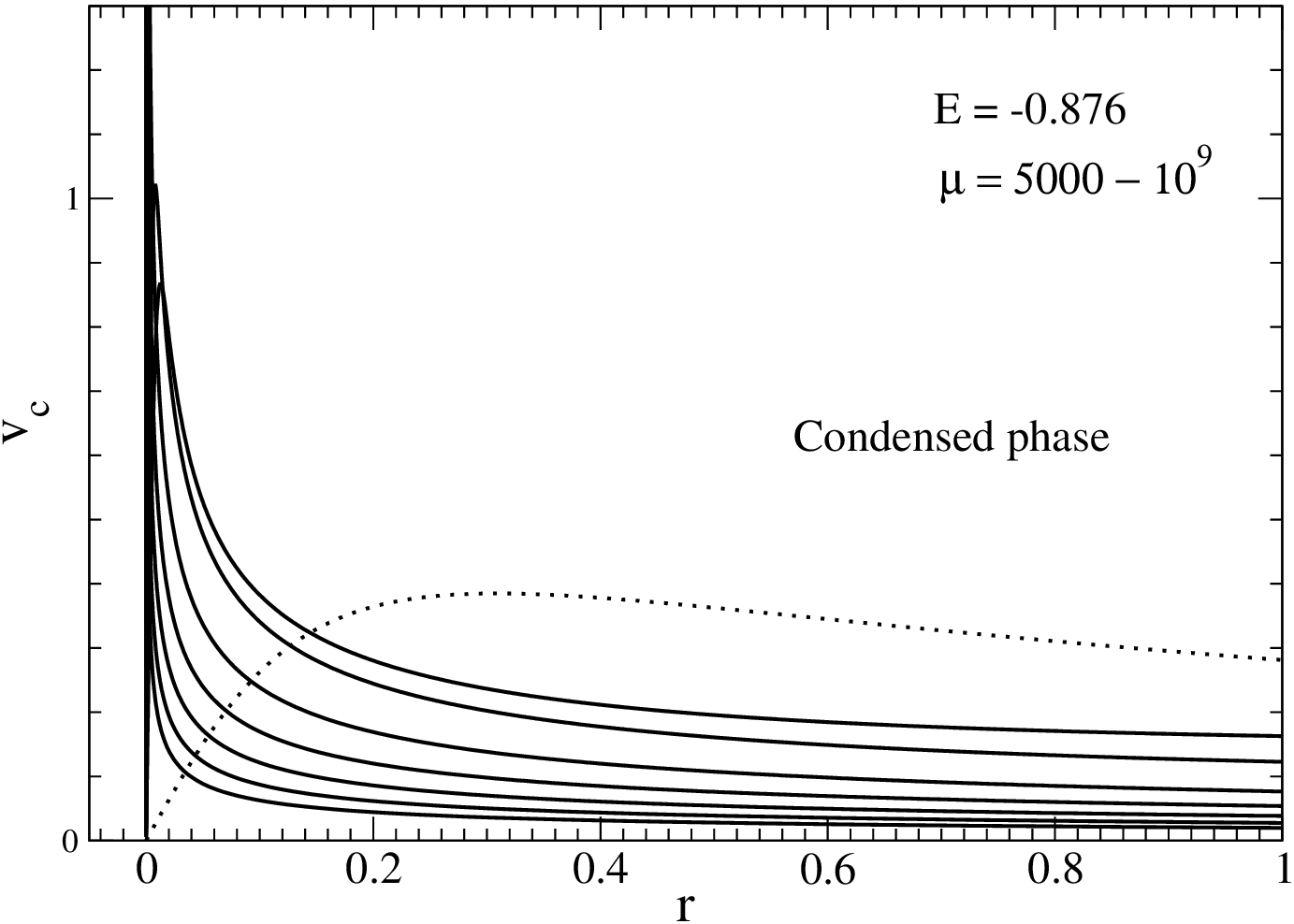}
\caption{Circular velocity of the condensed phase  (solution C) for different values of $\mu$ in linear
scales.}
\label{condensedVlindot}
\end{center}
\end{figure}

The solution A (gaseous phase) does not significantly change with $\mu$ and tends to the
classical King distribution for $\mu\rightarrow +\infty$. The density profile
and the rotation curve of the classical King model are represented as dotted
lines in Figs. \ref{embryonRHOdot}-\ref{condensedVlindot}.
Since the classical King model close to $E_c$ describes large dark matter halos relatively well (see Sec. \ref{sec_comp}) we shall
take it as
a reference in our discussion.

The solution B (embryonic phase) is  similar to the solution A (gaseous phase) except that it contains a
small embryonic nucleus.  Therefore, the solution B has a nucleus-halo
structure. The mass, the size and the absolute value of the potential energy of the nucleus
decrease as $\mu$ increases. As a result, for large $\mu$, the solutions A and B have
almost the same temperature ($\beta_A\simeq \beta_B$) and  the profiles A and B coincide
outside of the nucleus (see Figs. \ref{embryonRHOdot}-\ref{embryonVlindot}). This is why the branches A and B in the series of  equilibria superimpose
for $\mu\rightarrow +\infty$ (see Fig. \ref{mutresgrandprolongehenon}).  Still, the two solutions A and B are physically
distinct. In particular, the solution B is unstable as further discussed in Sec.
\ref{sec_canwe}.

\begin{figure}
\begin{center}
\includegraphics[clip,scale=0.3]{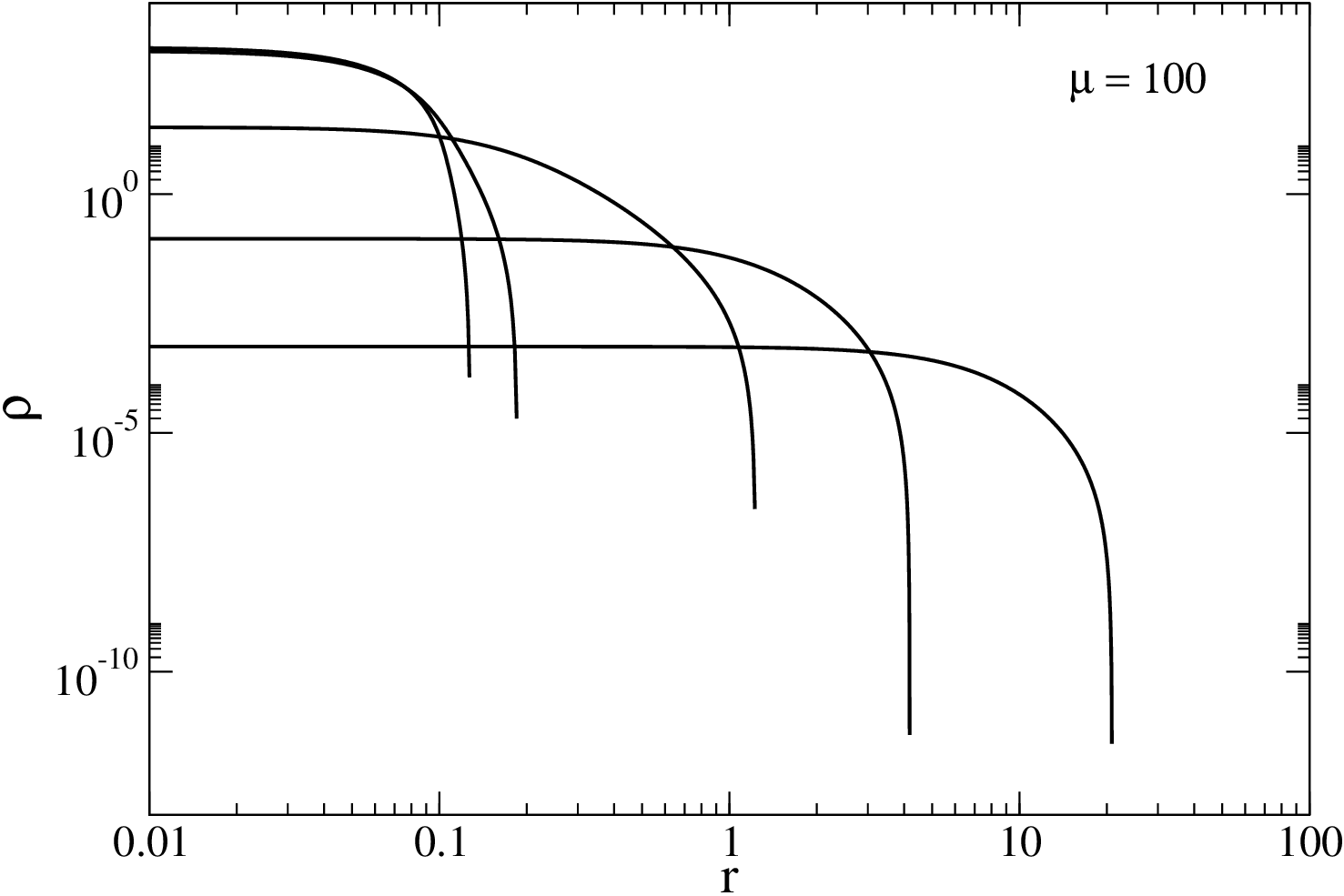}
\caption{Density profile along the series of equilibria for $\mu=100$ in
logarithmic scales. We have selected $k=0.142$ ($E=-0.0329$, $\beta=0.993$),
$k=1.31$ ($E=-0.181$, $\beta=1.66$) , $k=4.99$ ($E=-0.914$, $\beta=1.03$),
$k=18.0$ ($E=-4.27$, $\beta=1.03$),
and $k=41$ ($E=-4.65$, $\beta=2.51$).}
\label{densityMu100}
\end{center}
\end{figure}

\begin{figure}
\begin{center}
\includegraphics[clip,scale=0.3]{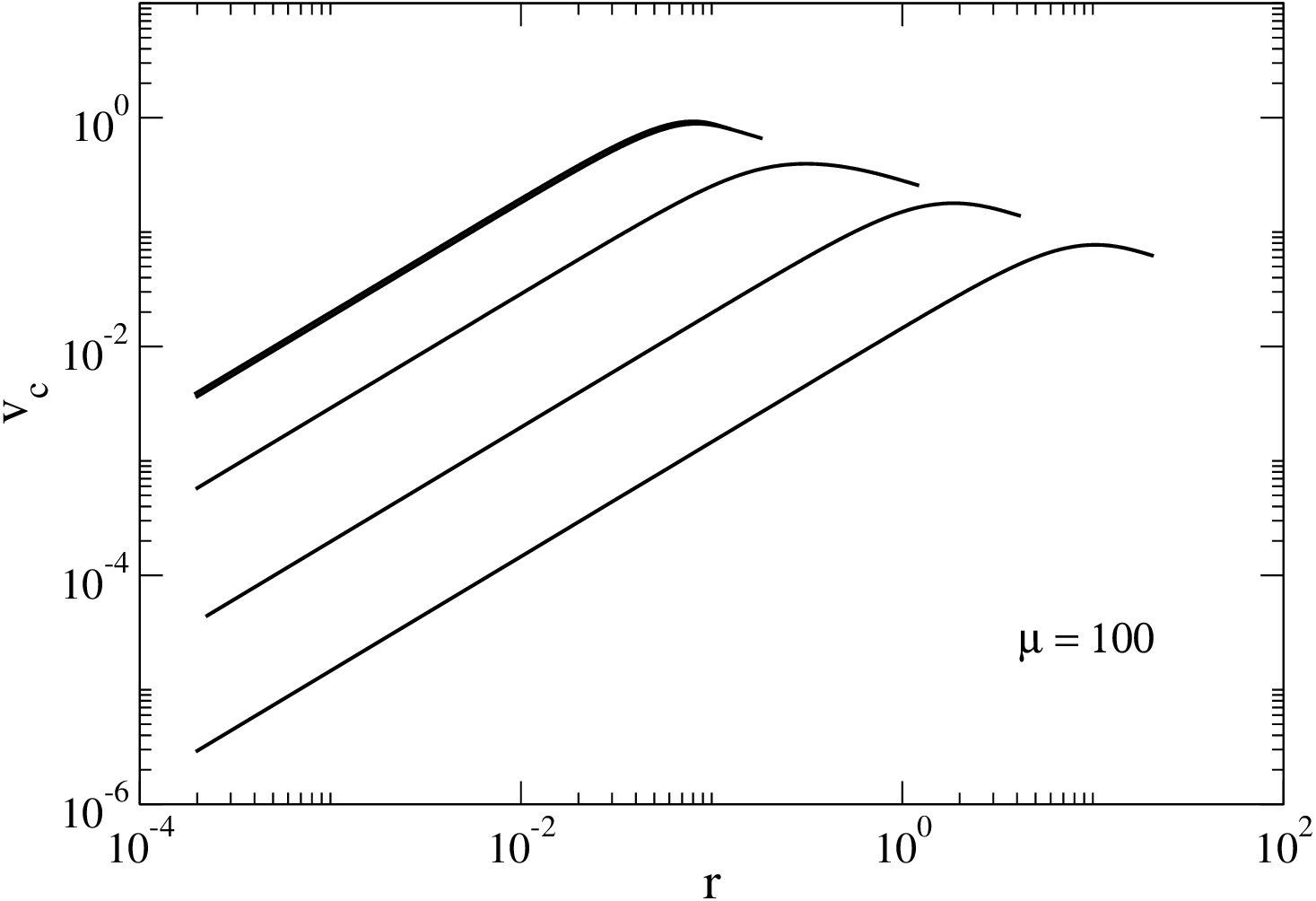}
\caption{Circular velocity along the series of equilibria for $\mu=100$ in
logarithmic scales. We have selected $k=0.142$, $k=1.31$, $k=4.99$, $k=18.0$,
and $k=41$.}
\label{velocityMu100}
\end{center}
\end{figure}

The solution C (condensed phase) is very different from the solution A (gaseous phase) and from the
solution B (embryonic phase). Like solution B, it has a nucleus-halo
structure. It contains
a small degenerate nucleus that has a small mass and a small radius. However, unlike solution B
it has a very negative potential
energy. As a result, the halo must be very hot in order to conserve the
total energy. This is why $\beta_C$ is small
(see Fig. \ref{mutresgrandprolongehenon}). Since the halo
is hot, it expands at very large distances (see Figs. \ref{condensedRHOdot}-\ref{condensedVlindot}). The halo
radius increases as $\mu$ increases. For $\mu\rightarrow
+\infty$, the mass and the radius of the
nucleus tend to zero but its potential energy tends to $-\infty$. As a result,
the temperature of the halo
tends to $+\infty$ and its radius also tends to $+\infty$. We therefore
obtain a  singular structure similar to
the ``binary $+$ hot halo'' mentioned in the Introduction (see footnote 3).
For finite $\mu$, quantum mechanics
provides a regularization of this singular structure: the ``binary'' is
replaced by a ``fermion ball'' whose size is fixed by quantum mechanics.

We now consider the effect of starting from the gaseous phase, and decreasing the
energy below the critical energy $E_c$. This is a natural evolution since
the energy $E(t)$ decreases, and the concentration parameter $k(t)$ increases,
as the system slowly evaporates \cite{cohn}. When $E<E_c$, the  system
undergoes a gravitational
collapse towards the condensed phase (solution C). According to the preceding discussion,
the result of the gravitational collapse is to form a degenerate nucleus
(fermion ball) of
much smaller mass than the initial cluster and to expel  a hot and massive  envelope
at very large
distances.\footnote{Things are very different in the canonical ensemble. For
$T<T_c$ the
system collapses and forms a degenerate nucleus containing most of the
mass. In that case, there
is almost no atmosphere. The core-halo structure is a property of the
microcanonical ensemble
that is not present in the canonical ensemble \cite{ijmpb}.} This is similar to
the formation of
red-giants and to the supernova explosion  phenomenon, except that it takes
considerably much more time (of the order of the Hubble time!) since the
gravothermal catastrophe is a rather slow process. At
the end, since the envelope is expelled, only the degenerate nucleus remains.
This could be a mechanism of formation of dwarf dark matter halos that are
completely degenerate.

We would like to point out that the previous results are valid
for large values of $\mu$ (large halos). For small
values of $\mu$ (small halos), and in particular for $\mu<\mu_{MCP}=1980$, the series of
equilibria unwinds and all the King profiles are stable. In that case, there is no phase transition (no collapse) anymore. The King profiles in the region of negative specific heat have a core-halo structure but the distinction
between the degenerate core and the
halo is not clear-cut (see Figs. \ref{densityMu100} and \ref{velocityMu100}).
These solutions may describe dwarf and intermediate size halos where degeneracy effects
are important.

\section{Can large dark matter halos harbor a fermion ball?}
\label{sec_canwe}

Many observations have revealed that galaxies and dark matter halos contain a
very massive object at the center. This compact object is usually interpreted as
a black hole. Alternatively, some authors have suggested that this object could
actually be a fermion ball made of the same matter as the rest of the halo.
Indeed, some configurations of the self-gravitating Fermi gas at finite
temperature have a nucleus-halo structure resembling a large dark matter
halo with a small compact object at the center. This nucleus-halo structure is
particularly clear in the embryonic phase (solution B). These solutions are
similar to the gaseous phase (solution A) except that they contain a small
degenerate nucleus. The halo is similar to a truncated classical isothermal gas
consistent with the observations of large dark matter halos (Burkert profile)
and the nucleus has the form of a degenerate fermion ball. When $\mu$ is large,
the fermion ball is very small so it does not affect the structure of the halo.
The corresponding density profiles and rotation curves are represented in Figs.
\ref{embryonRHOdot}-\ref{embryonVlindot}. The nucleus creates  a secondary peak
and a dip in  the rotation curve at very small radii that may not be resolved
observationally.  This type of nucleus-halo
configurations has been obtained by several authors \cite{gao,csmnras,viollier,ijmpb}. Some of them \cite{viollier} made the interesting suggestion that the fermion ball
could mimic the effect of a central black hole. However, these authors
\cite{viollier} did not investigate the stability of such configurations. Our
study (see also \cite{csmnras,pt,ijmpb}) shows that these structures (solution
B) are
thermodynamically unstable ({\it i.e.} unreachable)  because they are saddle
points of entropy at fixed mass and energy. Therefore, large dark matter halos
should not contain a degenerate nucleus (fermion ball). This is an important
prediction of
our study. The fact that fermion balls are not observed at the center of
galaxies (a central black hole is indeed observationally favored over a fermion ball
\cite{nature,reid}) is in agreement with our result.

We note that the solutions of the condensed phase (solution C) also have a core-halo structure with a degenerate nucleus and a non degenerate envelope. These solutions are stable. However, in that case, the nucleus formed by gravitational collapse releases an enormous energy that heats the envelope and disperse it at very large distances. As a result, only the degenerate object remains at the end. These solutions do not resemble a large dark matter halo with a central nucleus because the atmosphere is too hot (compare solutions B and C in Figs. \ref{embryonRHOdot}-\ref{condensedVlindot}). However, the nucleus alone resembles a dwarf halo that is a completely degenerate object without atmosphere.

\section{Can large dark matter halos harbor a black hole?}

We have seen in the previous section that the presence of a fermion ball at the 
center of large dark matter halos is unlikely because these nucleus-halo
structures are unreachable: they are saddle points of entropy. The presence of a
central black hole is more likely \cite{nature,reid}. These black holes could be
formed by the mechanism discussed by Balberg {\it et al.} \cite{balberg} if dark
matter is collisional. In that case, large dark matter halos may undergo a
gravothermal catastrophe when $E<E_{c}$. The increase of the density and
temperature of the core
during the collapse can trigger a dynamical (Vlasov) instability of general
relativistic origin leading to the formation of a central black hole. During
this process, only the core collapses. This can form a black hole of large
mass without affecting the structure of the halo. Therefore, this process leads
to large halos compatible with the Burkert profile for $r>0$ (see Sec.
\ref{sec_comp}) but harboring a central black hole at $r=0$.

In this scenario, the presence of black holes at the center of 
dark matter halos is conditioned by the possibility that dark matter halos may
undergo a gravothermal catastrophe. Now, when quantum mechanics is taken into
account, as in the fermionic King  model, an important result of our study is
the existence of a microcanonical critical point $\mu_{MCP}=1980$ below which
the
microcanonical phase transition (gravothermal catastrophe) is suppressed.
Roughly speaking, this result implies that ``large'' dark matter halos
($\mu>\mu_{MCP}$) that are non degenerate can undergo a gravothermal catastrophe
(although this is not compulsory\footnote{It is possible that a proportion of
large dark matter halos have a concentration parameter $k<k_{MCE}$ and have not
undergone core collapse (these halos do not contain a black hole) while some
halos have reached the critical threshold $k=k_{MCE}$ and have undergone core
collapse (these halos contain a black hole).}) and contain a central black hole 
while ``small'' dark matter halos ($\mu<\mu_{MCP}$) that are quantum objects
stabilized by the Pauli exclusion principle cannot contain a central black
hole because they do not experience a gravothermal catastrophe. This result
seems to qualitatively agree with the observations.

\section{A scenario of formation of dark matter halos}
\label{sec_scen}

We sketch below a general scenario of formation of dark matter halos
assuming that they are made of fermions such as massive neutrinos.

Initially, dark matter can be considered as  a spatially
homogeneous gas described by the relativistic Fermi distribution
$f=\eta_0^{Pauli}/(1+e^{pc/k_B T})$ where $\eta_0^{Pauli}=gm^4/h^3$ is the
Pauli bound \cite{tg}. The maximum value of the distribution
function is $f_0=(1/2)\eta_0^{Pauli}=(g/2)m^4/h^3$. Since this gas is
collisionless, it is described by the Vlasov-Poisson system. A spatially
homogeneous distribution is unstable and undergoes gravitational collapse 
(Jeans instability). The fermionic Jeans wavenumber $k_J=\sqrt{12\pi
G}(8\pi/3)^{1/3}m^{4/3}\rho^{1/6}/h$ \cite{cml} is finite so that quantum
mechanics
prevents the formation of small-scale structures and fixes a ground state. This
produces a sharp cut-off in the power spectrum.\footnote{In the CDM model where
$c_s^2=p'(\rho)=k_B T/m\rightarrow 0$, the classical Jeans wavenumber
$k_J=\sqrt{4\pi
G\rho}/c_s\rightarrow +\infty$. Therefore, if this model were valid, the
spatially homogeneous gas
would be unstable at all wavelengths and, consequently, structures would form at
all scales. Since there is no ground state, we would observe dark
matter halos of all sizes. The fact that we do not observe halos below a certain
scale (missing satellite problem) shows that quantum mechanics must be taken
into account in dark matter. Another possibility is to consider
warm dark matter (WDM) with $T\neq 0$. In that case, the Jeans wavenumber $k_J$
and the maximum value of the distribution function $f_0$ are determined by
thermal effects ({\it i.e.} by the velocity dispersion of the particles).} In
the
linear regime, some regions of over-density form. When the density has
sufficiently grown, these regions collapse under their own gravity at first in
free fall. Then, as nonlinear gravitational effects become important at higher
densities, these regions undergo damped oscillations (due to an exchange of
kinetic and potential energy) and finally settle into a quasi stationary state
(QSS) on a coarse-grained scale. This corresponds to the process of violent
relaxation first reported by Lynden-Bell \cite{lb} for stellar systems like
elliptical galaxies. This process is related to phase mixing and nonlinear
Landau damping. It is applied here to dark matter. In this context, the QSSs
represent dark
matter halos. Because of violent relaxation, the halos are almost isothermal and
have a core-halo structure.  The density of the core is relatively large and can
reach values at which quantum effects or Lynden-Bell's type of degeneracy are
important.\footnote{In the case of dark matter, the Lynden-Bell bound and the
Pauli bound are of the same order, differing by a factor two,
since $\eta_0^{LB}=f_0\sim \eta_0^{Pauli}/2=(g/2)m^4/h^3$.} On the other hand,
the halo is relatively hot and behaves more or less as a classical isothermal
gas. Actually, it cannot be exactly isothermal otherwise it would have an
infinite mass. The finite extension of the halo may be due to incomplete violent
relaxation \cite{lb}. The extension of the halo may also be limited by tidal
effects. In that case, the complete configuration of the system can be described
by the fermionic King model  \cite{mnras}.
As we have demonstrated, the fermionic
King model can show a wide diversity of configurations with different degrees of
nuclear concentration. The system can be everywhere non degenerate, everywhere
completely degenerate, or have a core-halo structure with a degenerate core and
a non degenerate halo. Small halos, that are compact, are degenerate. Their flat
core is due to quantum mechanics. Assuming that the smallest and most compact
observed dark matter halo of mass $M_h=0.39\, 10^6\, M_{\odot}$ and
radius $r_h=33\, {\rm pc}$ (Willman 1) is completely degenerate ($T=0$) leads to
a fermion mass of the
order of $1.23\, {\rm keV}/c^2$ \cite{vega,cml}. These particles may be sterile
neutrinos.  Small halos can merge with each other to form larger halos. This
is called hierarchical clustering. The merging of the halos also
corresponds to
a process of collisionless violent relaxation. Large halos, that are dilute, are
non degenerate. Their flat core is due to thermal effects.\footnote{Here, the
temperature is effective and it must be understood in the sense of Lynden-Bell.}
Knowing the mass of the fermions, we can deduce from the observations that halos
of
mass $0.39\, 10^6\, M_{\odot}<M_h<2.97\, 10^6 \, M_{\odot}$  are quantum
(degenerate) objects while halos of mass $M_h>2.97\, 10^6 \, M_{\odot}$  are
classical (non degenerate) objects \cite{vega,cml}. In the classical limit,
numerical simulations of violent relaxation generically lead to configurations
presenting an isothermal core and a halo whose density decreases as
$r^{-\alpha}$ with  $\alpha=4$ \cite{henonVR,albada,roy,joyce}. These
configurations are relatively close to H\'enon's isochrone profile. They can be
explained by models of incomplete violent relaxation
\cite{bertin1,bertin2,hjorth}. A density slope $\alpha=4$ is also consistent
with a King profile of concentration $k\sim 5$ \cite{cml}. If the halos were
truly collisionless, they would remain in a virialized configuration. However,
if the core is dense enough, collisional effects can come into play and induce
an evolution of the system on a long timescale (driven by
the gradient of temperature -velocity dispersion- between the core and the
halo) during which the concentration
parameter $k(t)$ increases while the slope $\alpha(t)$ of the density profile
decreases much like in globular clusters \cite{cohn}. We now have to distinguish
between small halos of mass $M_h<1.60\, 10^7\, M_{\odot}$ (corresponding to
$\mu<\mu_{MCP}=1980$) and large halos of mass $M_h>1.60\, 10^7\, M_{\odot}$
(corresponding to $\mu>\mu_{MCP}$) \cite{cml}. For small halos, the series of
equilibria (see Fig. \ref{ETmicro3}) does not present any instability so that
$k(t)$ increases and $\alpha(t)$ decreases regularly due to collisions and
evaporation. These halos are degenerate. They are stabilized against
gravitational collapse by quantum mechanics. As a result, they do not experience
the gravothermal catastrophe so they should not contain black holes. For large
halos, the series of equilibria (see Fig. \ref{ETmicro5summary}) presents an
instability at $k_{MCE}=7.44$. Because of collisions and evaporation, the
concentration parameter increases from $k\sim 5$ corresponding to a density
slope  $\alpha=4$ (a typical outcome of violent relaxation) up to the critical
value $k_{MCE}=7.44$  corresponding to a density slope $\alpha\sim 3$. Less
steep
halos ($\alpha<3$) are unstable ($k>k_{MCE}$). Large halos are
expected to be close to the point of marginal stability (see solution A in Fig.
\ref{profilesREAL}). At that point, the King profile can be approximated by the
modified Hubble profile that is relatively close to the Burkert profile fitting
observational halos. Some halos may be stable ($k<k_{MCE}$) but some halos may
undergo a gravothermal catastrophe ($k>k_{MCE}$). In that case, they
experience core collapse. The evolution is self-similar. The system develops
an isothermal core surrounded by a halo with a density slope $\alpha=2.2$
\cite{cohn,balberg}. The core radius decreases with time while
the central density and
the central temperature increase. The halo does not change. The
specific heat of the core is negative. Therefore, by loosing heat to the profit
of the halo, the core grows hotter and enhances the gradient of temperature
with the halo so the collapse continues. This is the origin of the gravothermal
catastrophe \cite{lbw}. For weakly collisional classical systems (globular
clusters), core collapse leads to a finite time singularity with
a density profile $\rho\propto r^{-2.2}$ at $t=t_{coll}$. The
singularity has infinite density but contains no mass. It corresponds to a
tight binary surrounded by a hot halo \cite{cohn}.  However, for
collisional dark matter halos, the situation is different. If
the particles are fermions, and if the mass of the halo is not
too large ($\mu>\mu_{MCP}$ not too large), the gravothermal catastrophe stops
when the core of the system becomes degenerate. This leads to a configuration
with a small degenerate nucleus (condensed state) surrounded by an extended
atmosphere that is relatively different from the structure of the halo before
collapse (see solution C in Figs. \ref{profilesREAL} and
\ref{condensedRHOdot}). However, the formation of this equilibrium
 structure can be very long (of the order of the Hubble
time) so that, on an intermediate timescale, the system is made of a
contracting fermion ball surrounded by an atmosphere that is not too much
affected by the collapse of the nucleus. Alternatively, if the halo mass is
large ($\mu>\mu_{MCP}$ large), during the gravothermal catastrophe the system
can develop a (Vlasov) dynamical instability of general relativistic origin and
form a central black hole without affecting the structure of the halo
\cite{balberg}.
In this way, the system is similar to the halo before collapse (Burkert profile)
except that it contains a central black hole.\footnote{More
precisely, the core collapse of fermionic dark matter halos is a two-stages
process.
In a first stage \cite{balberg}, the core collapses while the halo does not
change. Only the
density, the radius and the temperature of the core change. This creates strong
gradients of temperature between the core and the halo. At sufficiently high
temperatures (achievable  if $\mu$ is large) the system becomes relativistic and
triggers a dynamical instability leading to a black hole with a large mass.
Alternatively, if $\mu$ is small, quantum mechanics can stop the increase of the
central density and central temperature before the system enters in the
relativistic regime. In that case, core collapse stops. Then, in a second stage
(never studied until now because it requires quantum simulations),
the temperature uniformizes between the core and the halo. Therefore, the halo
heats up and extends at large distances until an equilibrium state with a
uniform temperature $T$ is reached (see solution C in Figs. \ref{profilesREAL}
and
\ref{condensedRHOdot}).} Large halos should not contain a
fermion ball because these nucleus-halos structures (see solution B in Figs.
\ref{profilesREAL} and \ref{embryonRHOdot}) are unreachable (saddle point of
entropy).

\section{On the collapse of large dark matter halos}

In this section, we come back on the different scenarios concerning the
possible collapse of large dark matter halos (when $\mu>\mu_{MCP}=1980$ and
$E<E_c$) and on
their resulting structure.

One possibility is that the collapse below $E_c$ leads to a core-halo
configuration with a dense and compact degenerate core (fermion ball), similar
to a white dwarf star at $T=0$, surrounded by a hot atmosphere. Actually, the
atmosphere
is so hot that it has the tendency to be expelled at large distances. This is
reminiscent of the red-giant phase where a star, having exhausted its nuclear
fuel, collapses into a white dwarf star and ejects its outer layers by forming a
planetary nebula. This is also reminiscent of the supernovae
explosion phenomenon \cite{supernovae} leading to a degenerate compact object
such as a neutron star or a black hole and to the expulsion
of a massive envelope. We may wonder whether a similar scenario can take place
(or has already taken place!) at the galactic scale. We may speculate
that many large dark matter halos are close to stable classical King models with
$k<k_{MCE}$ but that some
halos may reach the critical value $k=k_{MCE}$ and collapse to give birth to
degenerate dwarf dark matter halos of much smaller mass, with the expulsion of a
massive envelope. We emphasize, however, that this phenomenon takes
considerably much more
time  (of the order of the Hubble time) than the supernova phenomenon (a
few seconds) since
the gravothermal catastrophe is a rather slow process.

Another possibility is that the collapse below $E_c$ leads to the formation of a
black hole 
before quantum mechanics comes into play. In that case, we predict \cite{cml}
that large
dark matter halos with mass $M_h>1.60\, 10^7\, M_{\odot}$ may contain a central
black hole because they are non degenerate and can undergo a gravothermal
catastrophe ($\mu>\mu_{MCP}=1980$) while intermediate size and dwarf dark matter
halos with mass $M_h<1.60\, 10^7\, M_{\odot}$ should not contain a central black
hole because they are stabilized by quantum mechanics and cannot undergo a
gravothermal catastrophe ($\mu<\mu_{MCP}=1980$).

Finally, one could imagine that the collapse of dark matter halos leads to a fermion ball that could mimic the effect of a central black hole. However, one important conclusion of our study is that large
dark matter halos should {\it not} contain a degenerate nucleus (fermion ball)
because these nucleus-halo
configurations are thermodynamically unstable (saddle points of entropy). Therefore, they should be unreachable.

\section{Conclusion}

In this contribution, we have described some thermodynamical properties of the fermionic King
model introduced in \cite{stella,mnras}. A more detailed study is presented in \cite{cml}.
The interest of this distribution function with respect to
statistical mechanics was pointed out in \cite{ijmpb}: (i) this distribution has a finite mass, so
there is no need to enclose the system within an artificial  box; (ii) due to
the exclusion constraint, there
exist an equilibrium state for all accessible energies; (iii) this model exhibits interesting phase
transitions between gaseous and condensed states similar to those described in \cite{ijmpb} for a gas of self-gravitating
fermions enclosed within a box.

The fermionic King model also provides a realistic model of dark matter halos.
Dwarf dark matter halos of mass $M_h=0.39\, 10^6\, M_{\odot}$  are quantum
objects. They are completely degenerate and represent the ground state of the
sequence of dark matter halos. This fixes the mass of the fermions to about
$1.23\,
{\rm keV}/c^2$. Intermediate size halos of mass $0.39\, 10^6\,
M_{\odot}<M_h<2.97\, 10^6 \, M_{\odot}$ are partially degenerate. Large dark
matter halos of mass $M_h>2.97\, 10^6 \, M_{\odot}$ are non-degenerate so the
classical King model can be used. Many large halos seem to be 
close to the limit of marginal stability in MCE ($k\sim k_{MCE}$). The
corresponding  King profile
can be approximated by the modified Hubble profile. It has a flat core and a
halo in which the density decreases as $r^{-3}$.
The modified Hubble  profile is
relatively close to the
empirical Burkert profile that fits several rotation curves of galaxies. Some
halos may have undergone the gravothermal catastrophe ($k>k_{MCE}$). Two
possibilities can
happen. Core collapse may be stopped by quantum mechanics. In that case, the
system forms a degenerate object and expels  a hot and massive envelope. This
could be a mechanism responsible for the
formation of dwarf halos that are completely degenerate. Another possibility is
that the collapse triggers a dynamical instability of general relativistic
origin. In that case, the system forms a central black hole without affecting
the halo. This could be the mechanism responsible for the presence of black
holes at the center of galaxies. In order to form a black hole, the halo mass
must be sufficiently large ($M_h>1.60\, 10^7\, M_{\odot}$) so that the
gravothermal catastrophe can take place ($\mu>\mu_{MCP}$). Small halos
($M_h<1.60\, 10^7\,
M_{\odot}$) should not contain black holes because they do not experience the
gravothermal catastrophe ($\mu<\mu_{MCP}$). Therefore, the presence (or absence)
of
black holes at the center of galaxies may be connected to the existence
of a microcanonical critical point ($\mu_{MCP}=1980$) in the fermionic King
model \cite{cml}.
Finally, we have
shown that the presence of fermion balls at the center of dark matter halos is
unlikely because these nucleus-halo structures are unreachable (saddle
points of entropy). This may explain why black holes at the center of galaxies
are observationally favored over fermion balls \cite{nature,reid}.

Obviously, several configurations of dark matter halos  are possible within 
the fermionic King model making the study of this model  very rich. The system
can be non degenerate (large halos), partially degenerate (intermediate size
halos), or completely degenerate (dwarf halos). We can obtain core-halo
configurations with a wide diversity of nuclear concentration depending on $\mu$
({\it i.e.} the size of the system) and $E$. This may account for the
diversity of
dark matter halos observed in the universe. Large dark matter halos are non
degenerate classical objects. They may contain a black hole.
Small halos are degenerate quantum objects. They should not contain a
black hole. Our approach is the first attempt to determine the caloric curves of
dark matter halos. This allows us to study the thermodynamical stability of the
different configurations and to reject those that are unstable. In particular,
we have shown that the nucleus-halo configurations considered in the past (as in
Fig. \ref{embryonRHOdot}) are unstable. More work is needed to relate our
theoretical results to the observations.

\end{document}